\documentclass[
aip,
 amsmath,amssymb,
preprint,%
 reprint,%
]{revtex4-1}

\usepackage{graphicx}
\usepackage{xcolor}
\usepackage{dcolumn}
\usepackage{bm}
\usepackage[mathlines]{lineno}
\usepackage[utf8]{inputenc}
\usepackage[T1]{fontenc}
\usepackage{mathptmx}
\usepackage{etoolbox}
\makeatletter
\def\@email#1#2{%
 \endgroup
 \patchcmd{\titleblock@produce}
  {\frontmatter@RRAPformat}
  {\frontmatter@RRAPformat{\produce@RRAP{*#1\href{mailto:#2}{#2}}}\frontmatter@RRAPformat}
  {}{}
}%
\makeatother
\begin{document}

\preprint{AIP/123-QED}
\title{Chiral and Clock phases in Twisted Dipolar Clusters}
\author{Paula Mellado}
\email{paula.mellado@uai.cl}
\affiliation{Facultad de Ingeniería y Ciencias, Universidad Adolfo Ibáñez.\\  Diagonal las Torres 2640, Santiago, Chile.
}
\author{Xavier Cazor}
\affiliation{Facultad de Ingeniería y Ciencias, Universidad Adolfo Ibáñez.\\  Diagonal las Torres 2640, Santiago, Chile.
}
\author{Andres Concha}
\affiliation{Facultad de Ingeniería y Ciencias, Universidad Adolfo Ibáñez.\\  Diagonal las Torres 2640, Santiago, Chile.
}
\affiliation{Condensed Matter i-Lab, Universidad Adolfo Ibáñez, Diagonal las Torres 2640, Building D, Peñalolen, Santiago, Chile.}
\affiliation{CIIBEC, Research Center, Santiago, Chile.}
\affiliation{Theoretical Sciences Visiting Program, OIST Graduate University, 1919-1 Tancha, Onna, 904-0495, Japan.} 

\date{\today}
\begin{abstract}
We study samples and a dipolar model of magnetic rods arranged on twisted polygonal clusters in terms of the twist angle. We find that the relative twist between polygons induces noncollinear chiral phases, ranging from flux vortex closure to hedgehog like radial configurations.  
Chirality, quantified in terms of a bond order parameter, is an emergent property that behaves here as an Ising variable. The chiral configurations of the systems can be understood in terms of chirality and clock index order parameters, whose evolution with twist occurs through discontinuous switching of the magnetic textures. Within a fixed Ising chiral sector, the clock index, rooted in the $C_N$ invariance of the polygons, distinguishes chiral textures that share chirality.  As the twist increases, it continuously shifts the preferred relative clock phase, but the N-fold anisotropy only allows discrete orientations; the competition produces a tilted N-fold energy landscape whose global minimum hops discontinuously between clock sectors. As the number of sites in the polygon grows, the resulting response displays a nonlinear crossover from rigid, Ising-like behavior to an almost $\rm U(1)$-invariant regime, governed by a twist-induced suppression of the emergent $Z_N$ clock anisotropy. Guided by symmetry considerations and the outcomes of the numerical minimization, we developed a Landau phenomenological description that is compatible with both the Ising-type chirality and the $Z_N$ clock anisotropy.
\end{abstract}
\maketitle

\begin{quotation}
We investigate twisted bilayers of magnetic dipoles arranged on regular polygonal clusters. Despite the vanishing net magnetization, the system exhibits a finite vector chirality associated with flux-closure spin textures. Guided by simple experiments and a dipolar-interaction model, we show that geometric twisting stabilizes noncollinear chiral magnetic phases—of both circulating and radial character. Furthermore, we demonstrate that increasing the number of lattice sites forming the polygonal clusters continuously promotes the twist-induced, effectively Ising-like response toward an $\mathrm{U}(1)$-symmetric behavior. This evolution is associated with a pronounced suppression of the emergent $Z_N$ clock anisotropy as the polygon size grows in radially frustrated configurations. A Landau-theory description captures these phenomena and generalizes to twisted honeycomb bilayers, whose low-energy physics maps onto a sine-Gordon model with twist-tunable domain-wall phases.
\end{quotation}

\begin{figure*}
    \centering
\includegraphics[width=\linewidth]{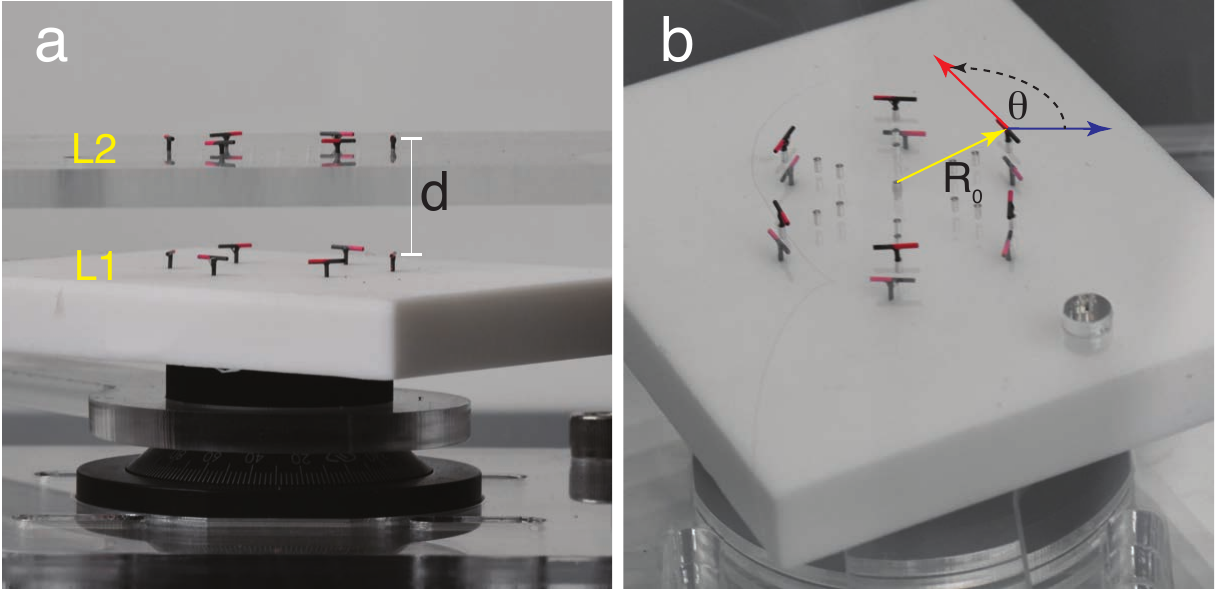}
\caption{\label{fig:f1} a.- Bilayer experimental setup: A $10$ mm thick Teflon plate (L1) holding $6$ Neodymium dipoles can rotate in the XY plane is mounted on a non-magnetic rotational stage. On top and parallel to this layer we set a second layer made out of $5$mm cast acrylic that is fixed to a vertical stage that allows fine control of the distance $d$. Both layers have $4$ mm depth holes that allow the graphite axes to freely rotate. b. Top-lateral view of the two layers for small twist $\phi$ between them. Top and bottom layer hexagons are concentric, an have a side of $R_{0}$. 
$\theta$ is the local rotation angle of a dipole respect to the x-axis (blue arrow).}
\end{figure*}

Chiral magnetic structures \cite{nagaosa2013topological,mohylna2022spontaneous} arise from broken inversion symmetry in the crystal unit cell \cite{camosi2018micromagnetics,pomeau1991three}. Magnetochiral textures are usually governed by antisymmetric, spin–orbit-driven interactions such as the Dzyaloshinskii–Moriya interaction (DMI) \cite{casher1974chiral,tomita2018metamaterials,thiaville1992twisted,dzyaloshinskii1960,dzyaloshinskii1958,togawa2016symmetry,nagaosa2013topological,mohylna2022spontaneous,camosi2018micromagnetics}. Yet DMI is not the only mechanism that stabilizes chiral textures; long-range dipolar interactions can also generate and stabilize chiral order in thin films and low-dimensional lattices \cite{lucassen2019tuning,mellado2023intrinsic,paula2023magnetic,yu2021chiral,ray2021hierarchy,shindou2013chiral,malozemoff2013magnetic,hubert2008magnetic,tapia2024chiral}. Magnetochirality has traditionally been engineered through the intrinsic crystal and electronic structure by material and composition choice~\cite{tokura2018nonreciprocal}. More recently, it has been shown that a small relative twist between bilayers with antiferromagnetic and ferromagnetic coupling can induce noncollinear spin configurations~\cite{kawamura2010chirality}, even without intrinsic chiral interactions~\cite{gong2017discovery,chen2013tailoring}.
\\
In lattices with discrete rotational symmetry, chirality is described by a clock-like angle \cite{elitzur1979phase}, whose energy minima correspond to the symmetry-allowed orientations of flux-closure textures \cite{sun2019phase}. Dipolar interactions favor such chiral states stabilizing vortical or radial configurations with finite vector chirality. Magnetochiral order then amounts to selecting one clock state, while fluctuations and defects appear as domain walls between distinct chiral sectors. Discrete-symmetry clock models \cite{jose1977renormalization,villain1975theory,cardy1996scaling} provide a minimal description of such systems where orientational order, interpolates between Ising-like and continuous $\rm U(1)$ behavior as the number of states increases \cite{elitzur1979phase}. They describe magnetic systems where lattice symmetry or anisotropic interactions pin the order parameter to discrete directions, producing domain walls, topological defects, and clock-anisotropy crossovers \cite{baek2013residual,nussinov2015compass}. 
\\
In the continuum limit, the $\rm Z_N$ clock model maps to a sine–Gordon field theory \cite{nussinov2015compass,kim2017partition}, where discrete anisotropy appears as a periodic locking potential for an angular order parameter \cite{gopinathan2004statistically}. Low-energy excitations are domain walls between neighboring clock states, with tension set by the clock pinning strength. In bilayers, twisting tunes this pinning: relative rotation frustrates the discrete anisotropy, softening the sine–Gordon potential and promoting domain-wall proliferation \cite{amit1980renormalisation,lemmens1986sine}. Twisted bilayers thus interpolate between regimes with isolated domain walls and dense domain-wall lattices, indicating an emergent restoration of $\rm U(1)$ symmetry at large twist or weak pinning.
\\
\emph{Summary of Results}
Aimed at understanding the role of geometrical twist in the fate of chiral magnetic phases in heterostructures, we study the ground states of a set of magnets located at the sites of a pair of polygons that are mutually rotated with respect to their centroid. Experiments with magnetic rods located at sites of twisted pairs of hexagons and exact minimization of the energy of  several twisted pairs of N-gons made out of point dipoles show the systems relaxing into noncollinear textures with finite intrinsic chirality. Depending on polygon geometry and twist, the systems host ordered chiral phases with circulating or inward–outward spin patterns. This chiral order arises from a twist-induced nonreciprocal interpolygon torque generated by the nonlocal magnetic field of the rotated layers. Acting as an effective chiral field, this torque reshapes the minimizing spin texture and produces sharp chirality switching with the twist angle. A Landau phenomenology captures these effects and extends to twisted honeycomb bilayers, whose low-energy theory maps onto a sine-Gordon model with twist-controlled domain-wall phases.
\\
We consider a pair of identical $N$-gons, each having $N$ easy-plane magnetic rods placed at its vertices, as shown in the samples made out of two hexagons shown in Fig.\ref{fig:f1}. The dipole moments are  $\bm m_i = m_0(\cos\theta_i, \sin\theta_i, 0)$, so the equilibrium configuration is fully specified by the set of
angles $\{\theta_i\}$ in each case. The two polygons are offset from each other by a constant distance d along the $z$ direction. The top polygon is rotated by a twist angle $\phi$ relative to the bottom one about the $z$ axis, as illustrated in Fig.~\ref{fig:f1}. The position of the $i$-th dipole is written as
\begin{equation}
\bm r_i(\phi)=
\begin{cases}
\bm R_i , & i=1,\dots,N,\\[4pt]
\mathcal R_z(\phi)\,\bm R_{i-N}+d\,\hat{\bm z},
& i=N+1,\dots,2N,
\end{cases}
\label{eq:positions}
\end{equation}
where $\bm R_i=R_N(\cos\frac{2\pi i}{N},\sin\frac{2\pi i}{N},0)$ denotes
the in-plane coordinates of a regular $N$-gon of unit radius, $R_N$ is the distance from the center of a dipole to the centroid of the polygon, d is the distance between polygons, and
$\mathcal R_z(\phi)$ is a rotation matrix about the $z$ axis. The distance between nearest neighbor dipoles is $a$.

\begin{figure*}
    \centering
\includegraphics[width=\linewidth]{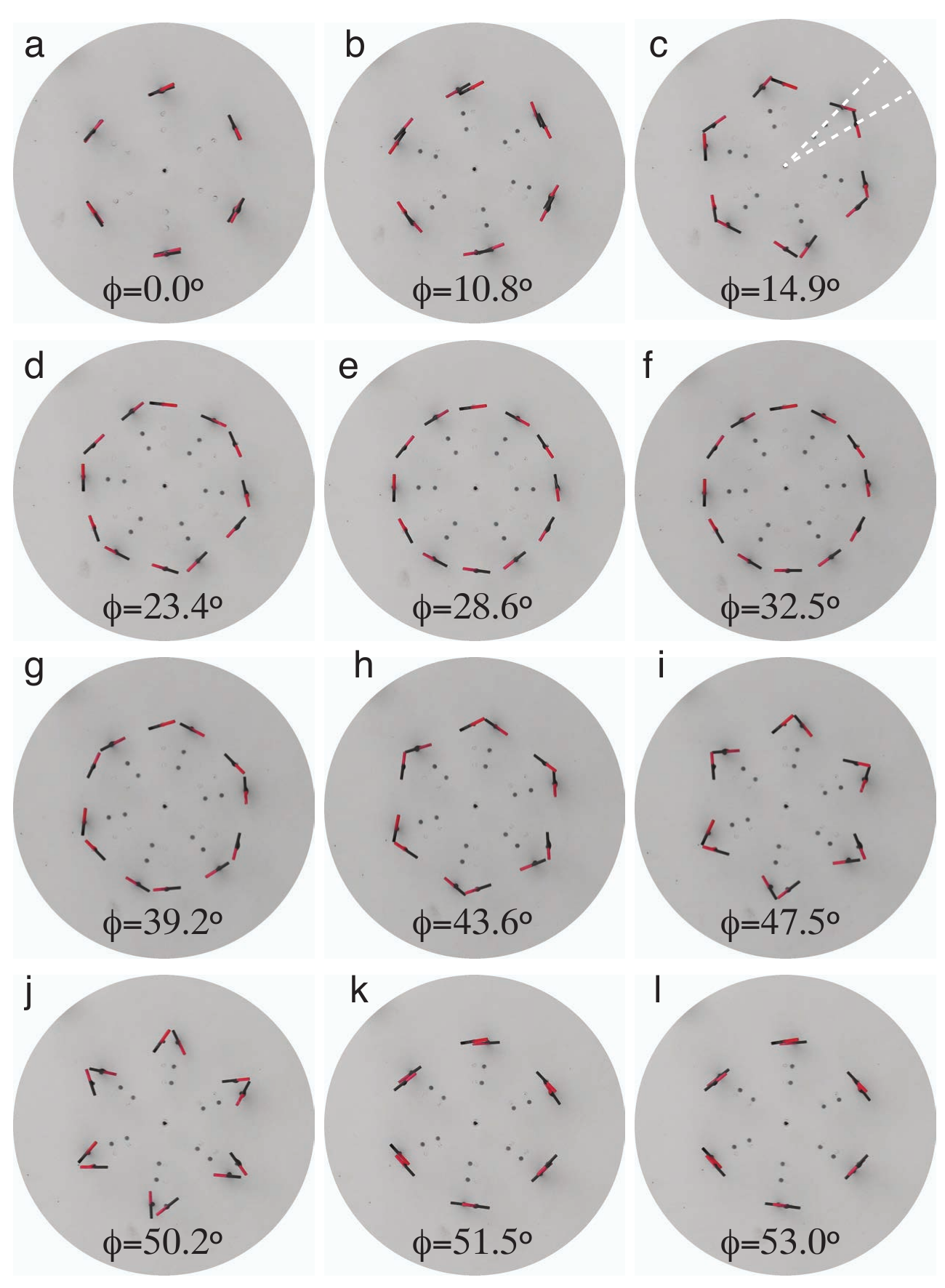}
\caption{\label{fig:f2} Evolution of the magnetic texture of the hexagonal samples as a function of the twisting angle $\phi$. The dashed white lines in Fig. 2-c indicate, as an example, a twisting angle of $\phi = 14.9^{\circ}$. In this configuration, the bottom hexagon is rotated while the upper (acrylic) hexagon remains fixed. The length scale of this panel is determined by the linear dimension of each magnet, $L = 5$ mm.}
\end{figure*}

Fig.~\ref{fig:f2} shows the relaxed state of experimental clusters of twisted hexagons (see Fig.\ref{fig:f1}) for twelve twist angles $\phi \in (0,\frac{2\pi}{6})$. At low twist ($\phi\approx0$), magnets form a closed flux (See Fig.~\ref{fig:f2}) in a head-to-tail arrangement. The magnetic rods in different polygons arrange antiferromagnetically. This configuration minimizes the total dipolar energy. As $\phi$ grows, the x-y symmetry breaks, and the twisted magnets rearrange in the x-y plane to preserve flux closure (Fig.\ref{fig:f2}(b-d)). For $\phi\approx 30^{\circ}$ (Fig.\ref{fig:f2}(e,f)), the top and bottom hexagonal sites are equidistant, and the magnets lie on a circle of radius $R_0$, with nearest neighbors belonging to different layers. For larger $\phi$, the process reverses (Fig.\ref{fig:f2}(g-l)). However, due to the lack of symmetry for internal fields, the system exhibits hysteresis\cite{concha2018designing}. Thus, configurations for $\phi$ increasing from $\pi/6$ to $\pi/3$ differ from those in the first half of the twist (Fig.\ref{fig:f2}(a-f)). All observed textures are chiral, but some states feature mostly antiparallel nearest neighbors, as in Figs.\ref{fig:f2}(i,j).

\begin{figure*}
\includegraphics[width=\linewidth]{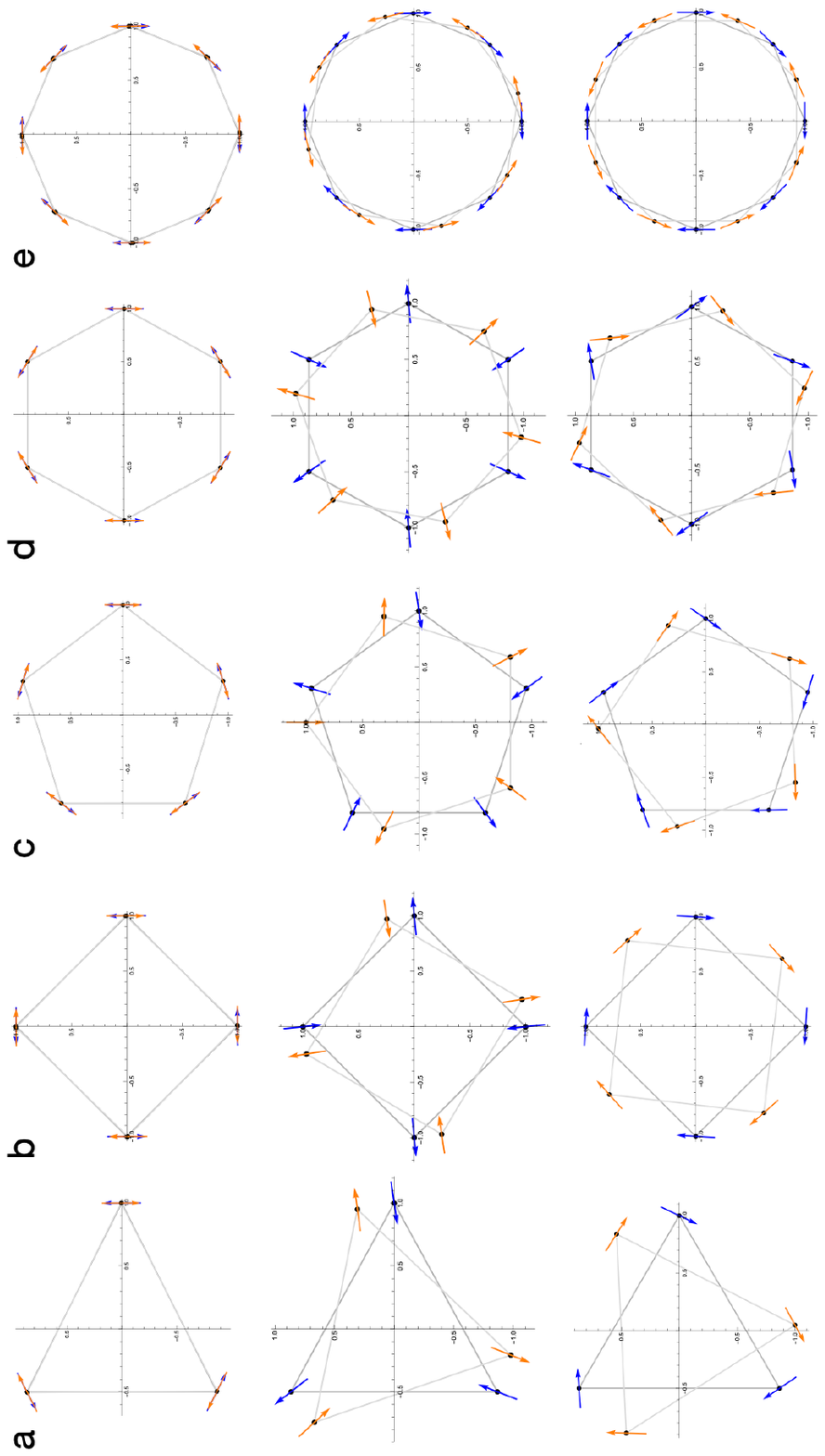}%
\caption{\label{fig:f3}Equilibrium magnetic configurations of  families of bilayers of polygons (N=3,4,5,6,8)  for (from top to bottom) a) $\phi=0$, $\phi=0.314$ and $\phi=0.572$, b) $\phi=0$, $\phi=0.249$ and $\phi=0.668$, c) $\phi=0$, $\phi=0.314$ and $\phi=0.36$, d) $\phi=0$, $\phi=0.328$ and $\phi=0.778$, e) $\phi=0$, $\phi=0.262$ and $\phi=0.393$. All angles are in radians}. Results are product of energy minimization of the dipolar Hamiltonian of the systems. 
\end{figure*}

\emph{Model} To investigate the origin of the experimentally observed textures, we represent the magnetic rods as point dipoles. Under this approximation, the total magnetostatic energy can be expressed as the sum of intrapolygon and interpolygon contributions:
\begin{equation}
H_{\mathrm{dip}}^{\mathrm{bil}}
=
H_{bb}+H_{tt}+H_{bt}(\phi),
\label{eq:Hbil_split}
\end{equation}
with
\begin{align}
H_{\ell\ell}
&=
\sum_{i<j\in \ell}^N
\left[
\frac{\bm m_{i\ell}\cdot\bm m_{j\ell}}{r_{ij,\ell\ell}^3}
-
3\frac{(\bm m_{i\ell}\cdot\bm r_{ij,\ell\ell})(\bm m_{j\ell}\cdot\bm r_{ij,\ell\ell})}{r_{ij,\ell\ell}^5}
\right],
\\
\nonumber
\ell\in\{b,t\},
\\
H_{bt}(\phi)
&=
\sum_{i\in b}^N\sum_{j\in t}^N
\left[
\frac{\bm m_{ib}\cdot\bm m_{jt}}{r_{ij,bt}^3}
-
3\frac{(\bm m_{ib}\cdot\bm r_{ij,bt})(\bm m_{jt}\cdot\bm r_{ij,bt})}{r_{ij,bt}^5}
\right],
\label{eq:H_inter}
\end{align}
$\{b,t\}$ labels the bottom and top polygons, $\bm r_{ij}=\bm{r}_i(\phi)-\bm{r}_j(\phi)$, and $m_0$ is the intensity of the magnetic moment. Hereafter, we use $m_0=1$, $a=1$, and $d=0.5$. The energy scale of our twisted clusters is the dipolar energy between nearest neighbor dipoles, $E_0=\frac{\mu_0m_0^2}{4\pi a^3}$. 
\\
For an angle $\phi$, the equilibrium magnetic texture is found by minimizing Eq.~\eqref{eq:Hbil_split} with respect to all angles $\{\theta_i\}$. Figs.~\ref{fig:f3}(a–e) show the equilibrium states for five polygonal families ($N=3,4,5,6,8$). At $\phi=0$, all clusters form an antiferromagnetic vortex state where top and bottom dipoles create a flux-closure texture. Neighboring polygons exhibit vortices with opposite handedness, which leads to a cancellation of the magnetic torques. We refer to such configurations as ferrochiral states. 
As $\phi$ increases, the in-plane components of the internal field develop an additional contribution proportional to $\phi$, inducing a torque along $\pm z$. This twist-induced torque reorients the magnets, giving rise to new ferrochiral textures that align with experimental observations (Figs.~\ref{fig:f2}(b–l)). 

For a single polygon, the z-component of the bond chirality \cite{cheong2022magnetic} is given by
\begin{equation}
\kappa =
\sum_{i=1}^{N}
\hat{\bm z}\cdot
\left(
\bm m_i\times\bm m_{i+1}
\right)
=
\sum_{i=1}^{N}
\sin(\theta_{i+1}-\theta_i),
\qquad \theta_{N+1}\equiv\theta_1 .
\label{eq:chirality_def}
\end{equation}
Bond chirality is defined identically for the top and bottom polygons, and $\kappa$ is even under time reversal. In the ferrochiral state, $\kappa_t=\kappa_b$, with $\kappa_{\ell} =\sum_{i=1}^{N}\sin(\theta_{i+1}^{(\ell)}-\theta_i^{(\ell)})$, with $\ell\in\{b,t\}$. For a regular polygon, the maximal value of $\kappa$ occurs when
$\theta_{i+1}-\theta_i = \Delta \theta=\frac{2\pi}{N}$,
\(
\kappa_{\max}(N)=\sum_{i=1}^{N}\sin\!\left(\frac{2\pi}{N}\right)
=
N\sin\!\left(\frac{2\pi}{N}\right).
\)
We computed bond chirality separately for the lower and upper polygons, $\kappa_{\rm b}$, $\kappa_{\rm t}$, and defined the cluster chirality as 
\begin{equation}
\chi=\frac{1}{2|\kappa_{\rm max}|}(\kappa_{\rm b}+\kappa_{\rm t})
\label{eq:chir}
\end{equation}

Depending on the sign of $\chi$, two topologically inequivalent chiral configurations can be distinguished. In the vortex state (V), the dipole moments within each polygon adopt a head-to-tail configuration, whereas the two polygons themselves are mutually aligned in an antiferromagnetic arrangement. In this case, $\kappa_{\rm \ell}>0$ and $\kappa_{\rm b}=\kappa_{\rm t}=+|\kappa_{\rm max}|$. In this case, the cluster chirality is $\chi=+1\equiv\chi_0$. Such magnetic textures are shown in Figs.\ref{fig:f2}(a-h), Figs.\ref{fig:f2}(k,l), and the top row of \ref{fig:f3}. The V state has nearly uniform angle differences $\Delta\theta\sim \pm 2\pi/N$ that favor dipolar interactions. 

In the radial (R) state, nearest-neighbor dipoles within each polygon assume antiparallel orientations, whereas the pair of polygons as a whole is arranged in a mutually antiferromagnetic configuration. Therefore, in this case one has $\kappa_{\rm \ell}<0$. When $N$ is even, each magnetic moment in a polygon is coupled antiferromagnetically to its two nearest neighbors, which leads to $\kappa=-|\kappa_{\rm max}|$, and consequently $\chi=-\chi_0$. In contrast, when $N$ is odd,  $-|\kappa_{\rm max}| < \kappa < 0$ and $-\chi_0 < \chi < 0$ (see Figs.~\ref{fig:f2}(i,j) and the middle row of Fig.~\ref{fig:f3}). This distinction originates from the fact that, for polygons containing an odd number of sites, an ideal perfectly regular R state cannot be realized, since at least one pair of rods is necessarily constrained to be parallel. To partially relieve this geometric frustration, the radial (hedgehog–antihog) configuration in such systems exhibits nonuniform angular separations between neighboring rods, which in turn reduces the magnitude of $\kappa$.

\begin{figure*}
\includegraphics[width=\linewidth]{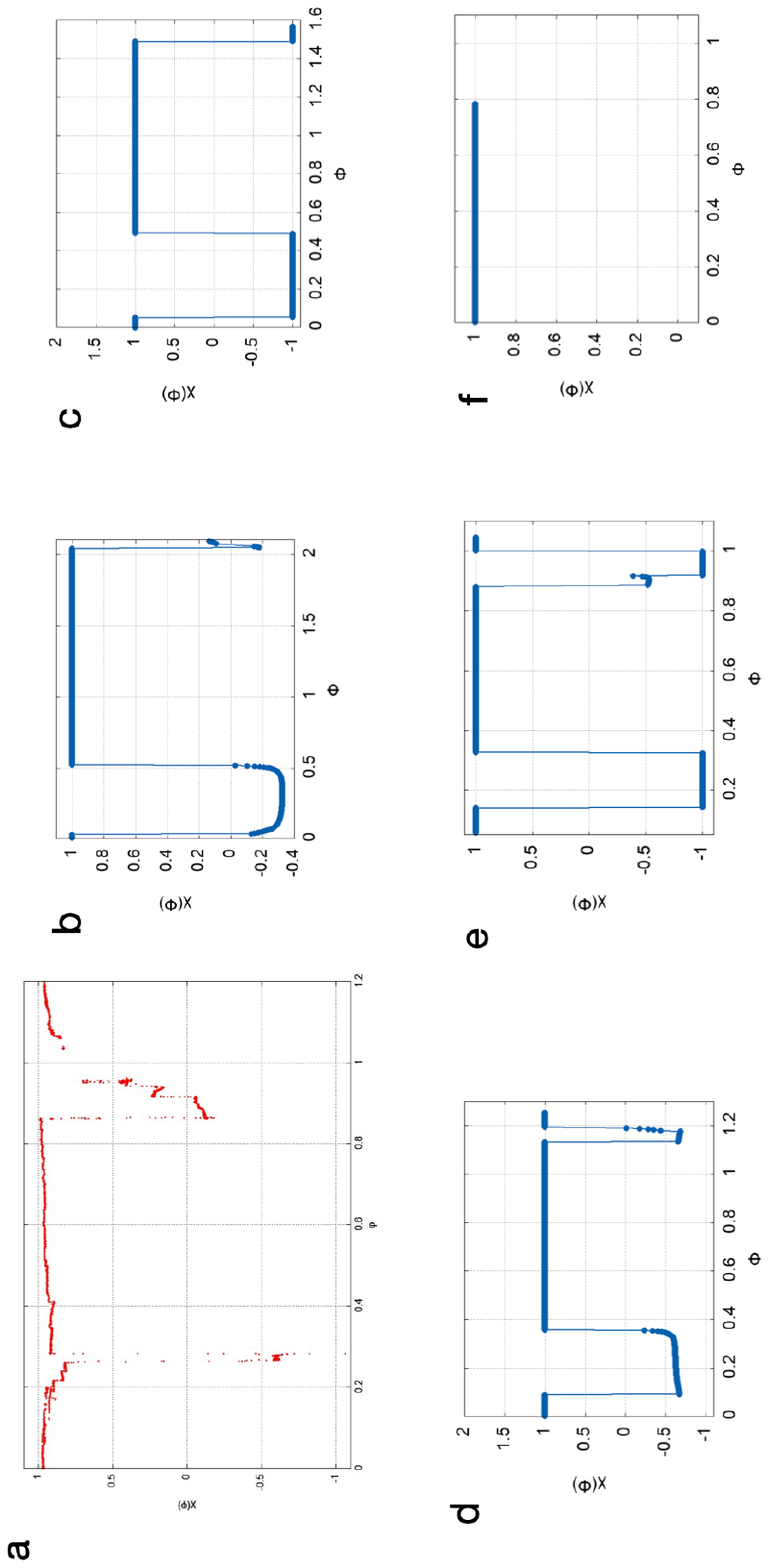}%
\caption{\label{fig:f4} a) Chirality $\chi(\phi)$ of experimentally realized twisted honeycomb samples as a function of the twist angle $\phi$. Panels b)–f) show $\chi(\phi)$ for the ground state of twisted dipolar clusters arranged in b) triangular, c) square, d) pentagonal, e) hexagonal, and f) octagonal polygonal geometries, respectively, as the twist angle was varied within the interval $\phi \in (0,2\pi/N)$. }
\end{figure*}

We observe that both the experimentally realized hexagonal twisted clusters and all numerically generated twisted N-gon configurations attain their minimum energy in a ferrochiral state \cite{batista2016frustration,yambe2023ferrochiral} for all values of $\phi$. We evaluated the cluster chirality as a function of the twist angle, $\chi(\phi)$, of our experimental samples (Figs.\ref{fig:f4}(a)), and for the minimum-energy configurations of twisted triangular, square, pentagonal, hexagonal, and octagonal clusters, considering twist angles in the interval $\phi\in(0,2\pi/N)$ (Figs.\ref{fig:f4}(b–f)). The experimentally extracted chirality follows the same qualitative trends and switching structure as the dipolar simulations, confirming that the observed configurations correspond to energy-minimizing states. For $N=3,4,5,6$ (Figs.\ref{fig:f4}(b–f)), $\chi$ depends strongly on $\phi$ and remains finite for all twist angles, displaying stepwise jumps as the system switches between topologically distinct textures. Because our bipolygonal structures only access V and R states, these configurations form two chiral Ising sectors that are topologically distinct and not connected by a smooth spin rotation. However, for the case $N\geq8$, Fig.\ref{fig:f4}(f), we found that $\chi$ is independent of $\phi$. 

Chirality curves for triangles, squares, pentagons, and hexagons collapse onto a common form (Fig.\ref{fig:f4}(b-f)), showing that the leading twist response is set by the $C_N$ clock anisotropy \cite{nussinov2015compass}, where  $\theta$ is restricted to N preferred directions related by the symmetry group C$_N$. Deviations from this collapse quantify higher-harmonic contributions with increasing $N$. This behavior demonstrates that the evolution from rigid, Ising-like chirality in the triangle, through multi-plateau clock behavior in the hexagon \cite{whitsitt2018quantum}, to an almost $U(1)$-invariant response in the octagon depicted in Fig.\ref{fig:f4} is governed by a clock-anisotropy crossover induced by the discrete polygonal symmetry under twisting.
We note that the angular susceptibility $\chi(\phi)$ of the samples shown in Fig.~\ref{fig:f4}(a) exhibits quantitative agreement with the twisted dipolar configuration with $N=6$ presented in Fig.~\ref{fig:f4}(f).

\emph{Landau functional of twisted clusters} Since the systems studied here are finite clusters, the observed discontinuities correspond to switching between competing local minima of the energy landscape rather than true thermodynamic phase transitions. Next, we use a Landau framework as an effective description of the energy landscape and its bifurcations under twist. The Landau functional is constructed on the basis of symmetry considerations and further constrained by insights obtained from numerical energy-minimization results.
At $\phi = 0$, the internal magnetic field $\bm B_{\phi=0}$ generated by the dipoles of one polygon is locally collinear with the dipole moments of the other polygon and, consequently, does not give rise to a torque. For finite values of $\phi$, the relative twist induces a transverse component of the magnetic field that is linear in $\phi$ at leading order, thereby producing a net torque directed along $\hat{\bm z}$.
Consider the effective energy density $F(\bm r)$ of one layer subject to the weak, slowly varying twist-induced magnetic field $\delta\bm B_{\rm{twist}}$. Within the linear response regime, it has been demonstrated previously \cite{mellado2023intrinsic} that performing a gradient expansion of the effective energy density \(F(\bm r)\) for a single layer subjected to a slowly varying, twist-induced magnetic field leads to the identification of the field conjugate to the chirality, \cite{mellado2023intrinsic}
\begin{equation}
h_{\kappa,N}
=
(\nabla\times\delta\bm B_{\mathrm{twist}})_z
\simeq
\phi\,(\partial_y B_{\phi=0}-\partial_x B_{\phi=0}),
\end{equation}
In the absence of $h_{\kappa,N}$, the system resides in one of two V-type configurations characterized by $\chi = +1$: a clockwise (CW) configuration or a counterclockwise (CCW) configuration. As the twist angle $\phi$ increases from zero, the system transitions into the R state, which introduces a new local minimum in the energy landscape. This local minimum is separated from the global minimum by an energy barrier whose height decreases monotonically with increasing $\phi$. At a critical twist $\phi = \phi^{c1}$, the V and R states become energetically degenerate. For $\phi > \phi^{c1}$, the R state becomes energetically favorable, and the system switches to the R configuration.

The sequence of magnetic ordering patterns exhibited by our twisted polygonal clusters as a function of $\phi$ can be rationalized by constructing a minimal Landau  effective energy functional for these systems. As an initial step in this analysis, we examine the dependence of $\kappa$ on the twisting angle $\phi$, when the system is in the ferrochiral state. In this case, the contribution of the Ising chiral order parameter to the effective energy takes the form,
$F_{\rm \kappa}(\kappa,\phi)
=\frac{r}{2}\kappa^2+\frac{u}{4}\kappa^4+\frac{w}{6}\kappa^6
-h_{\kappa,N}(\phi)\kappa,$
with $u<0,\; w>0$. At zero twist, the energy landscape exhibits a single minimum associated with the V state. As the absolute value of the twist angle $|\phi|$ increases, the term $h_{\kappa,N}(\phi)$ progressively tilts the potential, thereby stabilizing a second local minimum corresponding to the R configuration. This emergent minimum is separated from the primary one by an energy barrier whose height decreases monotonically with increasing $|\phi|$. The critical twist $\phi^{c1}$ can be estimated by employing
\[
\Delta F(\phi)
\equiv F_\chi(+\chi_0;\phi)-F_\chi(-\chi_0;\phi)
\simeq -2\,h_{\chi,N}(\phi)\,\chi_0.
\]
The discontinuous switching between the competing
minima takes place when the condition $h_{\chi,N}(\phi^{c1}) \simeq 0$ is satisfied. For the minimal symmetry-allowed functional form,
\[
h_{\chi,N}(\phi)=\lambda_\chi \sin(N\phi),
\]
this condition implies
\begin{equation}
\phi^{c1} \simeq \frac{\pi}{N},
\label{eq:phi_c1_estimate}
\end{equation}
In contrast to the triangular cluster, which exhibits only a single radial and a single vortex chiral phase, see Fig.~\ref{fig:f4}(b), the hexagonal twisted system displays multiple plateaus in $\chi$, see Fig.~\ref{fig:f4}(e). This behavior signals the presence of several distinct, clock-like chiral textures within the same Ising sector. This observation indicates the presence of lattice-harmonic effects that extend beyond the strictly rigid chiral regime associated with the triangular geometry. Although the dipolar Hamiltonian is, in principle, invariant under global spin rotations, this continuous symmetry is effectively reduced once the dipoles are constrained to occupy fixed lattice positions $\{\bm r_i\}$.

To investigate how the clock anisotropy evolves as the number of sites $N$ in a polygon increases, we consider low-energy chiral configurations characterized by angular variables $\{\theta_i\}$ together with a collective rotational degree of freedom $\theta_i \to \theta_i + \alpha$, where   $\alpha$ is a phase. Substituting it into the dipolar Hamiltonian gives a periodic energy $E(\alpha)$. Because the dipoles form a regular $N$-gon, the system is invariant only under $C_N$ rotations,
$E(\alpha)=E\!\left(\alpha+\frac{2\pi m}{N}\right)$, $m\in\mathbb Z$. The Fourier series of $E(\alpha)$ contains only $C_N$-compatible harmonics,
$E(\alpha)=E_0-\sum_{p=1}^{\infty} V_{pN}\cos(pN\alpha-\theta_{pN})$ where the leading term is the $N$-fold harmonic,
$E(\alpha)\simeq E_0-V_N\cos(N\alpha-\theta_0)$, yielding an emergent $Z_N$ clock anisotropy \cite{maryasin2016low}. The collective phase $\alpha$ is fixed by the internal angular correlations of the texture, where the clock-harmonic amplitude is set by the strength of dipolar correlations around the polygon. To leading order, one finds that $V_N \propto |\kappa|^N$, since the $N$-th harmonic arises from the coherent accumulation of angular phase differences along the closed polygonal loop, such that its amplitude is controlled by the product of local angular correlations. Consequently, clock anisotropy is pronounced in vortex-like configurations and is  reduced in radially symmetric textures. We identify the collective clock phase as $\vartheta \equiv \alpha$. Let $\bm r_{i\ell}(\phi)$ be the fixed position of site $i$ in polygon $\ell\in\{b,t\}$, with $\bm r_{ib}$ being independent of $\phi$, and $\bm r_{it}(\phi)$ obtained by rigidly rotating the top polygon. Each polygon has a collective clock phase $\vartheta_\ell$ associated with a collective rotation of the low-energy texture. Because each polygon has only $C_N$ symmetry, the intrapolygon energies generate the emergent clock pinning
\begin{equation}
F_{\mathrm{clock}}
=
-\sum_{\ell\in (t,b)}V_N^{(\ell)}|\kappa|^N\cos(N\vartheta_\ell-\theta_{0\ell})
\label{eq:F_clock_intra_bilayer}
\end{equation}
where $V_N$ quantifies the breaking of the $\rm U(1)$ rotational symmetry down to $C_N$. We define the $Z_N$-projected collective phase of polygon $\ell$ from the minimized microscopic angles $\{\theta_{i\ell}\}_{i=1}^{N}$.
A convenient definition is obtained from the $N$-fold harmonic
\begin{equation}
Q_{N,\ell}
\;\equiv\;
\frac{1}{N}\sum_{i=1}^{N} e^{\,iN\theta_{i\ell}},
\qquad
\vartheta_\ell
\;\equiv\;
\frac{1}{N}\arg\!\big(Q_{N,\ell}\big),
\label{eq:QN_theta}
\end{equation}
where $\vartheta_\ell$ is defined modulo $2\pi/N$.

\begin{figure*}
\includegraphics[width=\linewidth]{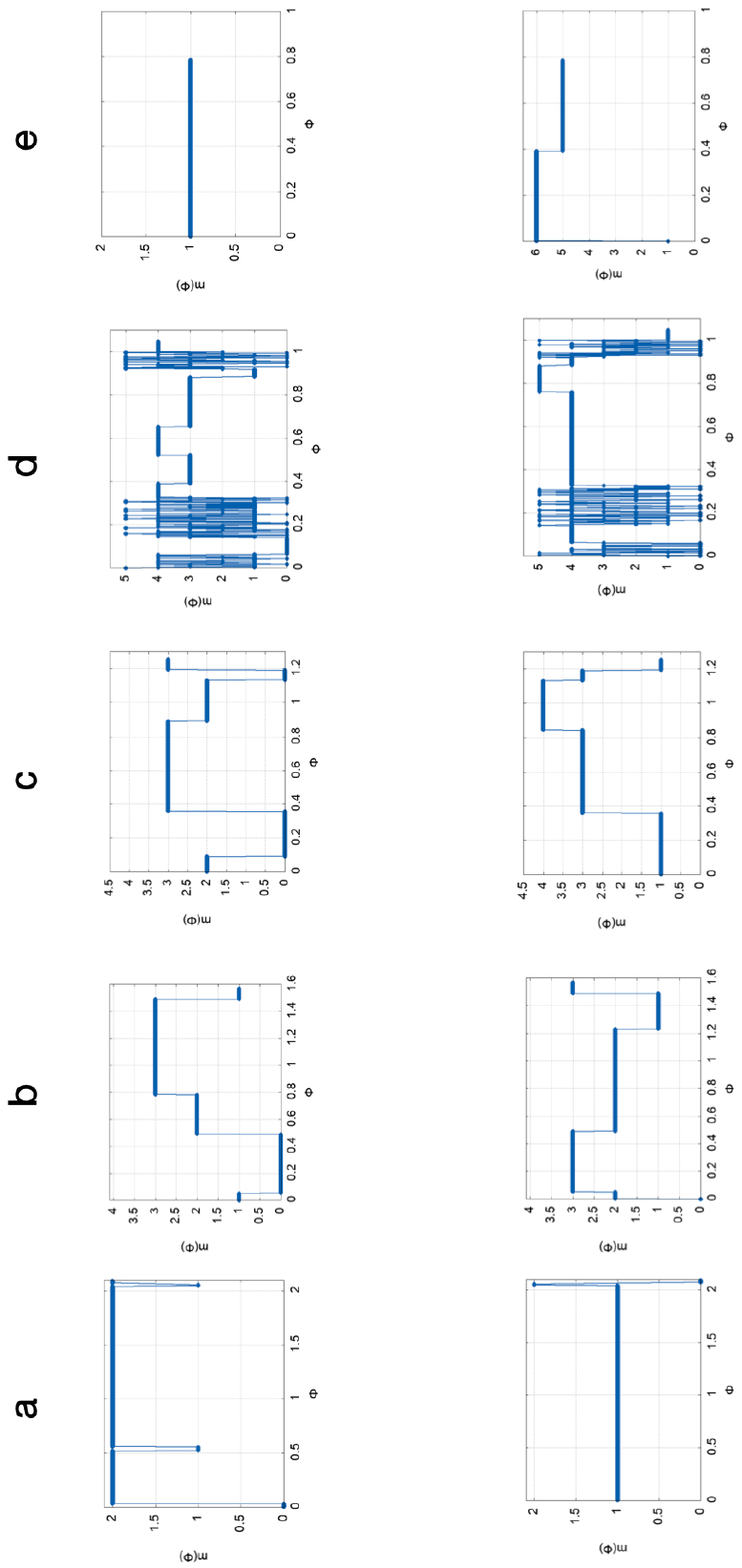}%
\caption{\label{fig:f5}Clock index $m(\phi)$ in the top and bottom polygon of twisted a) triangular, b) squared, c) pentagonal, d) hexagonal, and e) octagonal bipolygons as $\phi$ was tuned in the range $\phi\in(0,2\pi/N)$.}
\end{figure*}

The discrete clock index is obtained by binning $\vartheta_\ell(\phi)$ to the nearest $Z_N$ sector,
\begin{equation}
m_\ell(\phi)
\;=\;
\mathrm{Mod}\!\left(
\mathrm{Round}\!\left[
\frac{N}{2\pi}\big(\vartheta_\ell(\phi)-\vartheta_{0,\ell}\big)
\right],
\,N\right),
\label{eq:clock_index_def}
\end{equation}
where $\vartheta_{0,\ell}$ is a fixed gauge setting
$m_\ell=0$ for a chosen configuration at a selected twist. The inverse relation expresses the pinned clock phase in
terms of $m_\ell$ as
\begin{equation}
\vartheta_\ell(\phi)
\;\approx\;
\vartheta_{0,\ell}
+
\frac{2\pi}{N}\,m_\ell(\phi),
\qquad m_\ell\in\{0,1,\dots,N-1\},
\label{eq:theta_from_m}
\end{equation}
Fig.\ref{fig:f5}(a-e) shows $m(\phi)$ for $N=3,4,5,6,8$ respectively. The plateaus of $m_\ell(\phi)$ and their jumps by $\pm 1$ reflect discrete jumps between neighboring minima of the effective $Z_N$ clock potential with $\phi$. Between different clock textures, only the clock angle changes as they share the same chirality $\chi$, differing only by a global rotation, separated by clock-pinning barriers.  The $C_N$ pinning term in Eq.~\ref{eq:F_clock_intra_bilayer} arises from the $N$th angular Fourier harmonic of the underlying microscopic interaction energy. Its amplitude is suppressed with increasing $N$. As $N$ increases while the nearest-neighbor spacing is held fixed, the lowest-order symmetry-allowed harmonic is shifted to a higher order, and its amplitude becomes suppressed. At fixed inter-site spacing, increasing $N$ leads to a corresponding increase in the polygonal radius, $R_N \sim Na/(2\pi)$, such that the structure asymptotically approaches a continuous ring. In the case of the octagonal configuration, this limit suppresses the effective chiral field to zero, resulting in an approximately constant (flat) chiral response (Fig.~\ref{fig:f4}(f)). Examining Fig.\ref{fig:f5}(d) and Fig.\ref{fig:f4}(e), we find that the clock-phase switching is strongly enhanced in the R sector. This asymmetry stems from the different dipolar frustration patterns of the two chiral textures. In the R sector, the pronounced enhancement of clock-phase switching results from a strong suppression of the emergent $Z_N$ clock-pinning amplitude. In V textures, circulating spins add constructively to the $N$-fold harmonic, producing strong clock pinning and large energy barriers between clock sectors. In R textures, alternating angular differences cause destructive interference of the $N$-fold harmonic, strongly suppressing $V_N(\kappa)$. Consequently, the R sector has lower barriers and frequent clock-phase switching under twist.

Next, we investigate the contribution of the effect $\phi$ to the interpolygonal interactions. The dipolar interpolygonal coupling, which is governed by the characteristic dipolar energy scale, the interlayer separation, and the in-plane lattice constant, is described by
\begin{equation}
F_{\mathrm{inter}}
=
- J_\perp(\phi)\,|\kappa|^2\cos\!\big(\vartheta_t-\vartheta_b-\delta_0(\phi)\big).
\label{eq:F_lock_general}
\end{equation}
$J_\perp(\phi)$ denotes the effective interpolygonal stiffness, while $\delta_0(\phi)$ represents a geometric phase that characterizes the relative registry between the two polygons. For an $N$-gon, the clock phase $\vartheta$ has a period $2\pi/N$, so a uniform rotation by $\phi$ shifts the preferred interlayer phase difference by $N\phi$, $\delta_0(\phi)=\delta^* + N\phi$, 
where $\delta^*$ is a constant set by the untwisted stacking and the reference phases of $\vartheta_{t,b}$. $\delta_0(\phi)$ provides the leading twist dependence relevant for the clock-phase transitions. This interlayer contribution constitutes a coupling that constrains solely the relative phase between polygons, while leaving the overall (global) phase undetermined.

For polygons, the $\rm U(1)$ symmetry is reduced to $C_N$ by the leading lattice harmonic (clock anisotropy), which pins $\vartheta$ to one of $N$ discrete minima and breaks the gauge freedom, allowing for the chiral-field term $h_{N}(\phi)=\sum_m\lambda_{m,N}\sin(mN\phi)$. 
The twist-induced torque is odd under spatial inversion and therefore transforms as a pseudoscalar field, permitting a linear coupling exclusively to the bond chirality $\kappa$. This gives rise to a driving term of the form $F_{\mathrm{drive}}=-h_{\kappa,N}(\phi)\,\kappa$, which is contained in the chirality-dependent contribution $F_{\kappa}(\kappa,\phi)$. Within a fixed chiral sector, a secondary, symmetry-allowed leading contribution is the coupling to the clock degree of freedom, $-h_N(\phi)\cos(\vartheta-\vartheta_0)$, which energetically biases the selection of particular clock sectors while preserving the underlying $C_N$ symmetry. Thus, the twist plays the role of the conjugate field to the chirality, whereas its influence on $\vartheta$ manifests solely through periodic contributions within a given chiral sector. This geometric, Zeeman-like contribution therefore effectively acts as a field coupled to a single polygonal degree of freedom.

Overall, the slow collective modes of twisted polygonal clusters are described by the clock angle $\vartheta$ and the chirality $\kappa$. Increasing $N$ simultaneously (i) shifts the response to the higher harmonic and (ii) suppresses its amplitude, since both $V_N$ and $\lambda_{1,N}$ rapidly decrease as the polygon approaches a ring. This accounts for the progression in Fig.\ref{fig:f5}: for the triangle, the effective anisotropy is strong, and the dynamics reduce to Ising-like switching between two chiral sectors; for the hexagon, the clock term creates multiple metastable minima, and the chirality shows six plateaus corresponding to distinct clock-locked textures; for the octagon, the clock anisotropy and the leading harmonic of the twist-induced chiral drive are so weak that $\kappa(\phi)$ is flat.

Collecting the leading-order symmetry-permitted contributions described above, the coupled
chiral-clock effective energy functional for twisted polygons can be expressed in the form
\begin{widetext}
\begin{align}
F(\kappa,\vartheta_t,\vartheta_b;\phi)
&=
\frac{r}{2}\kappa^2
+\frac{u}{4}\kappa^4
+\frac{w}{6}\kappa^6
-
2h_{\kappa,N}(\phi)\,\kappa- V_N(\kappa)|\kappa|^N(\cos(N\vartheta_b-\theta_{0b})
- \cos(N\vartheta_t-\theta_{0t}))- \nonumber\\
&\quad J_\perp(\phi)\,|\kappa|^2\cos\!\big(\vartheta_t-\vartheta_b-\delta^*-N\phi\big),
\label{eq:Landau_bilayer_prelude}
\end{align}
\end{widetext}
(u<0, w>0). Eq.\ref{eq:Landau_bilayer_prelude} establishes a hierarchy of discontinuous switching between magnetic textures governed by a dominant chiral barrier $\sim w\kappa^N$ and a subdominant clock barrier $\sim V_N\kappa^N$. It predicts (i) an abrupt reversal of the system chirality, controlled by $\kappa$ and $h_{\kappa,N}(\phi)$, in which the twist-induced internal torque induces a discontinuous Ising-like jump of $\kappa$, while $2h_{\kappa,N}(\phi)$ tilts the $\kappa$ double-well potential; and (ii) for a fixed chiral sector, a cascade of twist-driven, clock-like discontinuous switchings of $\vartheta$, determined by the discrete $Z_N$ pinning and the phase mismatch $\delta_0(\phi)$.

The robustness of the predicted phases—plateau structure and the critical twist $\phi^{c1}$—depends sensitively on the interlayer spacing to lattice constant ratio, $d/a$. In our simulations, $d/a = 0.5$ corresponds to strong dipolar interlayer coupling (decaying as $1/r^3$). As $d/a$ increases, the twist-induced interpolygon torque and plateau width decrease, and $\phi^{c1}$ scales toward its single-layer limit. This trend matches experiments. Thus, moiré-induced chiral textures signal the strong-coupling regime and can be tuned or suppressed by varying the vertical separation between polygons.

\emph{Effective Hamiltonian} By employing the Landau phenomenological framework expressed in Eq.~\ref{eq:Landau_bilayer_prelude}, we derive the corresponding effective Hamiltonian governing the twisted clusters in a given chiral sector $\chi$. In this case, the chirality amplitude is fixed, we set $
\kappa \;\to\; \kappa_\chi$, $ (\chi~\text{fixed})$, 
and define
\begin{equation}
V_N^{(\chi)} \equiv V_N(\kappa_\chi)\,|\kappa_\chi|^N,
\qquad
K_\perp^{(\chi)}(\phi) \equiv J_\perp(\phi)\,|\kappa_\chi|^2 .
\end{equation}
Dropping constant factors of $\kappa$, the only remaining slow variables are the clock phases. The interlayer coupling fixes the relative phase, and when it is appreciable, the last term in the effective energy locks the relative angle and can be dropped
\(
\vartheta_t-\vartheta_b \equiv \vartheta_- \sim \delta_0(\phi)(\mathrm{mod}\;2\pi)
\), which allows the use of a single clock variable
\[
\vartheta \equiv \vartheta_b,
\qquad
\vartheta_t = \vartheta + \delta_0(\phi).
\]
which yields the effective angular Hamiltonian in the chiral sector $\chi$.
\begin{widetext}
\begin{align}
H^{(\chi)}(\vartheta;\phi)
&=
- V_N^{(\chi)}\cos\!\big(N\vartheta-\theta_{0b}\big)
- V_N^{(\chi)}\cos\!\Big(N\vartheta + N\delta_0(\phi)-\theta_{0t}\Big)
\;
\label{eq:H1cl_rawsum}
\end{align}
\end{widetext}
We define $A \equiv N\vartheta-\theta_{0b}$, and
$B(\phi) \equiv N\delta_0(\phi)-(\theta_{0t}-\theta_{0b})
$. In \eqref{eq:H1cl_rawsum}, this yields
\begin{equation}
H^{(\chi)}(\vartheta;\phi)=
- V_{\mathrm{eff}}^{(\chi)}(\phi)\,
\cos\!\big(N\vartheta-\theta_{\mathrm{eff}}(\phi)\big)
\label{eq:Hsingle_final}
\end{equation}
with
\begin{align}
V_{\mathrm{eff}}^{(\chi)}(\phi)
&\equiv
2V_N^{(\chi)}\left|\cos\!\Big(\frac{B(\phi)}{2}\Big)\right|,
\label{eq:Veff_def}
\\
\theta_{\mathrm{eff}}(\phi)
&\equiv
\theta_{0b}-\frac{B(\phi)}{2}
\quad
\Big(\mathrm{mod}\;2\pi\Big),
\label{eq:thetaeff_def}
\end{align}
\begin{align}
B(\phi)=N\delta^*+N^2\phi-(\theta_{0t}-\theta_{0b}),
\\
\nonumber
\theta_{\mathrm{eff}}(\phi)=\theta_{\mathrm{eff}}(0)-\frac{N^2}{2}\phi
\quad(\mathrm{mod}\;2\pi).
\label{eq:thetaeff_phi_linear}
\end{align}
The amplitude $V_N^{(\chi)}$ follows from the microscopic dipolar Hamiltonian after minimizing over the angles $\{\theta_i\}$ in each polygon. The twist alters the relative in-plane dipole alignment between polygons, thereby shifting the preferred phase and renormalizing the pinning scale \eqref{eq:Veff_def}. In a fixed chiral sector, strong interlayer dipolar locking removes the relative clock phase, leaving a single $Z_N$ clock variable $\vartheta$ with twist-induced phase drift $\theta_{\mathrm{eff}}(\phi)$ and a chirality-dependent pinning scale $V_{\mathrm{eff}}^{(\chi)}(\phi)$. For fixed $N$, the polygon geometry yields the lowest symmetry-allowed anisotropy term $\cos(N\vartheta)$ in the effective energy. 
\\
Plateaus, such as those depicted in Fig.\ref{fig:f5}, correspond to intervals during which $\vartheta(\phi)$ remains confined within a single $2\pi/N$ sector; a switching event is defined to occur when $\vartheta(\phi)$ crosses the boundary between adjacent sectors. The effective Hamiltonian \eqref{eq:Hsingle_final} has minima
\begin{equation}
\vartheta_k(\phi)=\frac{\theta_{\mathrm{eff}}(\phi)+2\pi k}{N},
\qquad
k\in\mathbb{Z}.
\label{eq:minima_branches}
\end{equation}
As $\phi$ increases, each minimum branch of $\theta_{\mathrm{eff}}(\phi)$ drifts. The observed integer plateaus arise because the index $m(\phi)$ 
is a sector label that remains constant until $\vartheta_k(\phi)$ crosses a sector boundary at
\begin{equation}
\vartheta=\vartheta_0+\left(m+\frac{1}{2}\right)\frac{2\pi}{N},
\qquad m\in\mathbb{Z}.
\end{equation}
A switch $m\to m\pm 1$ occurs at a twist angle $\phi_c$ where the clock variable satisfies
\begin{equation}
\vartheta_k(\phi_c)=\vartheta_0+\left(m+\frac{1}{2}\right)\frac{2\pi}{N}.
\label{eq:boundary_crossing}
\end{equation}
Using \eqref{eq:minima_branches}, 
\begin{equation}
\theta_{\mathrm{eff}}(\phi_c)
=
N\vartheta_0 + \left(m+\frac{1}{2}\right)2\pi - 2\pi k.
\label{eq:thetaeff_crossing}
\end{equation}
If $\theta_{\mathrm{eff}}(\phi)$ is monotonic and dominated by a linear drift, then the critical twists 
\begin{equation}
\phi_c(m;k)
=
\frac{2}{N^2}\left[
\theta_{\mathrm{eff}}(0)-N\vartheta_0
-\left(m+\frac{1}{2}\right)2\pi+2\pi k
\right]
\label{eq:phi_crit_explicit}
\end{equation}
This predicts an equally spaced plateau structure with spacing
\begin{equation}
\Delta\phi=
\phi_c(m+1;k)-\phi_c(m;k)=\frac{4\pi}{N^2}.
\label{eq:plateau_spacing}
\end{equation}

\emph{Outlook} In the following, we extend the phenomenological framework developed above to interpret previously reported results in twisted bilayer systems and to gain insight into the expected continuum limit \cite{hejazi2020noncollinear}. Consequently, we do not present explicit simulations of extended twisted lattices in this work and instead defer such an analysis to future studies.

Our analysis focuses on the twisted honeycomb lattice, in which each hexagonal plaquette is endowed with a $Z_6$ clock degree of freedom, represented by a corresponding phase variable. The introduction of a relative twist between the two layers generates a spatially varying phase mismatch in the interlayer coupling. Within this framework, we accordingly define the associated coarse-grained clock-phase fields
$\vartheta_\ell(\bm{r})=\frac{1}{6}\text{arg}\!\left(\frac{1}{6}\sum_{j\in hex(r,\ell)}e^{i6\theta_j}\right)$,
at the hexagonal plaquette centers $\bm{r}$. In the long-wavelength limit, and in direct analogy with Eq.\ref{eq:Landau_bilayer_prelude}, the minimal continuum functional consistent with the underlying symmetries is given by
\begin{widetext}
\begin{align}
\mathcal{F}
&=
\int d^2r\,
\Bigg\{
\frac{\rho_b}{2}\,|\nabla \vartheta_b|^2
+\frac{\rho_t}{2}\,|\nabla \vartheta_t|^2
- g_b \cos\!\big(N\vartheta_b-\theta_{0b}\big)
- g_t \cos\!\big(N\vartheta_t-\theta_{0t}\big)
- K_\perp(\phi)\cos\!\big(\vartheta_t-\vartheta_b-\delta_0(\bm{r};\phi)\big)
\Bigg\}
\label{eq:SG_bilayer_start}
\end{align}
\end{widetext}
$\rho_{b,t}$ are the intralayer phase stiffnesses, $g_{b,t}$ are the local single-hexagon $Z_6$ anisotropies, and $K_\perp(\phi)$ is the chirality-selective stiffness \cite{vignaud2023evidence} that locks the relative orientation of polygonal textures between layers.
It has been demonstrated in twisted bilayer systems \cite{hejazi2020noncollinear} that the phase mismatch acquires a spatially varying character, $\delta_0(\bm{r};\phi)=\delta^*+\bm{G}$ with $\bm{G}\sim 6{\bm q}_m(\phi)\cdot {\bm r}$ and $q_m$ the moiré lattice wavevector. In a lattice, $\delta_0$ becomes slowly varying since $
|\bm{G}(\phi)|\propto \phi$
We define
\[
\vartheta_+(\bm{r})=\frac{\vartheta_t+\vartheta_b}{2},
\qquad
\vartheta_-(\bm{r})=\vartheta_t-\vartheta_b.
\]
For large $K_\perp$, $\vartheta_-(\bm{r})$ is large and locks to
$\vartheta_-(\bm{r})\sim \delta_0(\bm{r};\phi)$ canceling the last term in Eq.\ref{eq:SG_bilayer_start}.
Substituting $\vartheta_t=\vartheta_++\delta_0/2$ $\vartheta_b=\vartheta_+-\delta_0/2$
gives an effective single-field functional
\begin{widetext}
\begin{align}
\mathcal{F}_{\mathrm{SG}}[\vartheta_+]
=
\int d^2r\,
\left[
\frac{\rho_{\mathrm{eff}}}{2}|\nabla \vartheta_+|^2
- g_{\mathrm{eff}}(\bm{r};\phi)\,
\cos\!\big(N\vartheta_+ - \Theta_{\mathrm{eff}}(\bm{r};\phi)\big)
\right],
\label{eq:SG_singlefield}
\end{align}
\end{widetext}
with an effective stiffness $\rho_{\mathrm{eff}}\simeq (\rho_b+\rho_t)/2$
and moiré-modulated respective amplitudes and phases
\begin{align}
g_{\mathrm{eff}}(\bm{r};\phi)\propto
\left|\cos\!\Big(\frac{N}{2}\delta_0(\bm{r};\phi)-\frac{\theta_{0t}-\theta_{0b}}{2}\Big)\right|,
\\
\Theta_{\mathrm{eff}}(\bm{r};\phi)=
\theta_{0b}-\frac{N}{2}\delta_0(\bm{r};\phi)+\frac{\theta_{0t}-\theta_{0b}}{2}.
\label{eq:SG_geff_Thetaeff}
\end{align}
Equation \eqref{eq:SG_singlefield} corresponds to the continuum formulation of the $Z_N$ clock sine-Gordon field theory \cite{cuevas2014sine,flamino2020lattice,coppersmith1983pinning,miller1986nonlinear}. For uniform $\delta_0(\phi)$ (single-cluster twist), $\Theta_{\mathrm{eff}}$ drifts uniformly, causing global clock-phase drift and discrete sector switching in $m(\phi)$. For spatially varying $\delta_0(\bm{r};\phi)$ in bilayer moiré systems, the effective phase field $\Theta_{\mathrm{eff}}(\bm{r};\phi)$ becomes a slowly varying function of position that induces the formation of solitons or domain walls, and realizes commensurate–incommensurate phenomena \cite{bruce1978theory,popov2011commensurate} of the Frenkel–Kontorova type \cite{lebedeva2019commensurate,sasaki1989symmetry} in the clock field $\vartheta_+(\bm{r})$. At small twist (strong pinning), Eq.\ref{eq:SG_geff_Thetaeff} yields a commensurate regime where $\vartheta_-$ is locked near $\vartheta_-\sim \frac{\theta_0}{6}+\frac{\pi}{3}m(r)$, so $m(r)$ forms domains. Domain walls  are lines with an energy cost $\sim \sqrt{\rho g}$. While for single clusters $\phi^{c1}$ marks the chiral discontinuous jumps, in a twisted bilayer, there are specific positions $\bm r^c$ where the local driving phase meets the switching condition $\delta_0(\bm{r}^c;\phi^c)\sim \frac{\pi}{3}(m+\frac{1}{2})$. Domain walls appear along these lines. At larger twists, $\delta_0(\bm{r};\phi)$ winds in space, so the bilayer cannot remain in a single clock sector. To accommodate the misfit, the system forms a network of phase slips where $\vartheta_-$ jumps by $\pi/3$, producing an incommensurate soliton network \cite{lebedeva2019energetics,wang2024electrically,kaliteevski2023twirling,cazeaux2023relaxation,park2025unconventional}.

\emph{Conclusions} To investigate the influence of geometrical twist on the stability and evolution of chiral textures in magnetic systems, we analyzed isolated clusters consisting of pairs of twisted polygons hosting magnetic moments at their vertices. Chirality, quantified by means of a bond-order parameter, emerges as a collective property that, in this context, effectively behaves as an Ising-like variable. The chiral configurations of the systems can be quantitatively described by two distinct order parameters, whose dependence on the twist angle exhibits a sequence of discontinuous jumps. Within a given Ising-chiral sector, the clock index provides a classification scheme for chiral textures that share the same sign of chirality. Whereas the twist couples to the total chirality, acting as a conjugate field that induces discontinuous switching between distinct chiral configurations, the clock phase itself exhibits additional discontinuous rearrangements arising from the competition between N-fold crystalline pinning and a twist-dependent interpolygon phase mismatch. As the twist angle is increased, the preferred relative clock phase is continuously displaced; however, the N-fold anisotropy constrains the system to a discrete set of orientations. The interplay between these effects generates a tilted N-fold energy landscape in which the global minimum undergoes discontinuous jumps between distinct clock sectors. As N becomes large, the leading symmetry-allowed harmonics in the clock potential are shifted to progressively higher orders.
A Landau functional captures the two distinct discontinuous switching processes and provides a starting point for deriving, within a given chiral sector, the effective Hamiltonian of the system. In this regime, strong interlayer dipolar locking removes the relative clock phase, reducing the clock variables of the two polygons to a single 
$Z_N$ clock variable. These results allowed us to extend the model to a twisted honeycomb bilayer, where the low energy theory maps onto a sine-Gordon model with twist controlled domain-wall phases.
\begin{acknowledgments}
P.M. and X.C. acknowledge support from the Fondo Nacional de Desarrollo Científico y Tecnológico (Fondecyt) under Grant No. 1250122. AC acknowledges partial support from FONDECYT (Chile) through Grant 1250681 and by the Theoretical Sciences Visiting Program (TSVP) at OIST Graduate University (Japan).
\end{acknowledgments}
\section*{Data Availability Statement}
The data that support the findings of this study are available from the corresponding author upon reasonable request.
\appendix
\section{Experimental Methods} The two hexagonal arrays were fabricated using acrylic and polytetrafluoroethylene (Teflon) plates. Each vertex (site) of the hexagons is designed to accommodate a single magnetic XY rod. In the samples, each rod is a cylindrical neodymium magnet with radius $r = 0.25 \times 10^{-3}\text{m}$, length $L = 5 \times 10^{-3}\text{m}$, mass $m = 0.007 \times 10^{-3}\text{kg}$, saturation magnetization $M_\mathrm{s} = 0.92 \times 10^{6}\text{A/m}$, and magnetic moment $m_0 = 9.0 \times 10^{-4}\text{A·m}^2$. The magnets were mounted vertically using graphite rods as low-friction pivot axes, constraining rotation to the horizontal plane, thereby implementing the easy-plane anisotropy condition required.

The experimental platform (Fig.\ref{fig:f1}) was mounted on a vibration-isolated optical table with active damping, thereby maintaining alignment stability throughout the rotational protocol. The lower lattice was mounted on a Newport RM-25A rotation mount, enabling controlled angular reorientation with sub-degree resolution. The upper lattice was supported by an Edmunds 281 Lab Jack, which served as a mechanically stable elevation base providing precise vertical positioning. Our measurements showed that magnetic coupling became appreciable around $h_c \approx 10 $mm inter-hexagon separation. For $h > h_c$, hexagons ordered independently. Throughout, the interlayer separation was fixed at $h = 8$ mm for all twist angles.

This experimental realization allowed us to directly observe equilibrium states and magnetic textures as the hexagonal layers rotated (Fig.\ref{fig:f2}). We recorded the process using a Nikon D850, extracting the rods angular positions at all twist angles using standard segmentation techniques.
 

\begin{thebibliography}{55}%
\makeatletter
\providecommand \@ifxundefined [1]{%
 \@ifx{#1\undefined}
}%
\providecommand \@ifnum [1]{%
 \ifnum #1\expandafter \@firstoftwo
 \else \expandafter \@secondoftwo
 \fi
}%
\providecommand \@ifx [1]{%
 \ifx #1\expandafter \@firstoftwo
 \else \expandafter \@secondoftwo
 \fi
}%
\providecommand \natexlab [1]{#1}%
\providecommand \enquote  [1]{``#1''}%
\providecommand \bibnamefont  [1]{#1}%
\providecommand \bibfnamefont [1]{#1}%
\providecommand \citenamefont [1]{#1}%
\providecommand \href@noop [0]{\@secondoftwo}%
\providecommand \href [0]{\begingroup \@sanitize@url \@href}%
\providecommand \@href[1]{\@@startlink{#1}\@@href}%
\providecommand \@@href[1]{\endgroup#1\@@endlink}%
\providecommand \@sanitize@url [0]{\catcode `\\12\catcode `\$12\catcode `\&12\catcode `\#12\catcode `\^12\catcode `\_12\catcode `\%12\relax}%
\providecommand \@@startlink[1]{}%
\providecommand \@@endlink[0]{}%
\providecommand \url  [0]{\begingroup\@sanitize@url \@url }%
\providecommand \@url [1]{\endgroup\@href {#1}{\urlprefix }}%
\providecommand \urlprefix  [0]{URL }%
\providecommand \Eprint [0]{\href }%
\providecommand \doibase [0]{http://dx.doi.org/}%
\providecommand \selectlanguage [0]{\@gobble}%
\providecommand \bibinfo  [0]{\@secondoftwo}%
\providecommand \bibfield  [0]{\@secondoftwo}%
\providecommand \translation [1]{[#1]}%
\providecommand \BibitemOpen [0]{}%
\providecommand \bibitemStop [0]{}%
\providecommand \bibitemNoStop [0]{.\EOS\space}%
\providecommand \EOS [0]{\spacefactor3000\relax}%
\providecommand \BibitemShut  [1]{\csname bibitem#1\endcsname}%
\let\auto@bib@innerbib\@empty
\bibitem [{\citenamefont {Nagaosa}\ and\ \citenamefont {Tokura}(2013)}]{nagaosa2013topological}%
  \BibitemOpen
  \bibfield  {author} {\bibinfo {author} {\bibfnamefont {N.}~\bibnamefont {Nagaosa}}\ and\ \bibinfo {author} {\bibfnamefont {Y.}~\bibnamefont {Tokura}},\ }\bibfield  {title} {\enquote {\bibinfo {title} {Topological properties and dynamics of magnetic skyrmions},}\ }\href@noop {} {\bibfield  {journal} {\bibinfo  {journal} {Nature nanotechnology}\ }\textbf {\bibinfo {volume} {8}},\ \bibinfo {pages} {899--911} (\bibinfo {year} {2013})}\BibitemShut {NoStop}%
\bibitem [{\citenamefont {Mohylna}\ \emph {et~al.}(2022)\citenamefont {Mohylna}, \citenamefont {Albarrac{\'\i}n}, \citenamefont {{\v{Z}}ukovi{\v{c}}},\ and\ \citenamefont {Rosales}}]{mohylna2022spontaneous}%
  \BibitemOpen
  \bibfield  {author} {\bibinfo {author} {\bibfnamefont {M.}~\bibnamefont {Mohylna}}, \bibinfo {author} {\bibfnamefont {F.~G.}\ \bibnamefont {Albarrac{\'\i}n}}, \bibinfo {author} {\bibfnamefont {M.}~\bibnamefont {{\v{Z}}ukovi{\v{c}}}}, \ and\ \bibinfo {author} {\bibfnamefont {H.}~\bibnamefont {Rosales}},\ }\bibfield  {title} {\enquote {\bibinfo {title} {Spontaneous antiferromagnetic skyrmion/antiskyrmion lattice and spiral spin-liquid states in the frustrated triangular lattice},}\ }\href@noop {} {\bibfield  {journal} {\bibinfo  {journal} {Physical Review B}\ }\textbf {\bibinfo {volume} {106}},\ \bibinfo {pages} {224406} (\bibinfo {year} {2022})}\BibitemShut {NoStop}%
\bibitem [{\citenamefont {Camosi}\ \emph {et~al.}(2018)\citenamefont {Camosi}, \citenamefont {Rougemaille}, \citenamefont {Fruchart}, \citenamefont {Vogel},\ and\ \citenamefont {Rohart}}]{camosi2018micromagnetics}%
  \BibitemOpen
  \bibfield  {author} {\bibinfo {author} {\bibfnamefont {L.}~\bibnamefont {Camosi}}, \bibinfo {author} {\bibfnamefont {N.}~\bibnamefont {Rougemaille}}, \bibinfo {author} {\bibfnamefont {O.}~\bibnamefont {Fruchart}}, \bibinfo {author} {\bibfnamefont {J.}~\bibnamefont {Vogel}}, \ and\ \bibinfo {author} {\bibfnamefont {S.}~\bibnamefont {Rohart}},\ }\bibfield  {title} {\enquote {\bibinfo {title} {Micromagnetics of antiskyrmions in ultrathin films},}\ }\href@noop {} {\bibfield  {journal} {\bibinfo  {journal} {Physical Review B}\ }\textbf {\bibinfo {volume} {97}},\ \bibinfo {pages} {134404} (\bibinfo {year} {2018})}\BibitemShut {NoStop}%
\bibitem [{\citenamefont {Pomeau}(1991)}]{pomeau1991three}%
  \BibitemOpen
  \bibfield  {author} {\bibinfo {author} {\bibfnamefont {Y.}~\bibnamefont {Pomeau}},\ }\bibfield  {title} {\enquote {\bibinfo {title} {Three short stories on chiral structures in condensed matter},}\ }in\ \href@noop {} {\emph {\bibinfo {booktitle} {Growth and Form: Nonlinear Aspects}}}\ (\bibinfo  {publisher} {Springer},\ \bibinfo {year} {1991})\ pp.\ \bibinfo {pages} {415--429}\BibitemShut {NoStop}%
\bibitem [{\citenamefont {Casher}\ and\ \citenamefont {Susskind}(1974)}]{casher1974chiral}%
  \BibitemOpen
  \bibfield  {author} {\bibinfo {author} {\bibfnamefont {A.}~\bibnamefont {Casher}}\ and\ \bibinfo {author} {\bibfnamefont {L.}~\bibnamefont {Susskind}},\ }\bibfield  {title} {\enquote {\bibinfo {title} {Chiral magnetism (or magnetohadrochironics)},}\ }\href@noop {} {\bibfield  {journal} {\bibinfo  {journal} {Physical Review D}\ }\textbf {\bibinfo {volume} {9}},\ \bibinfo {pages} {436} (\bibinfo {year} {1974})}\BibitemShut {NoStop}%
\bibitem [{\citenamefont {Tomita}\ \emph {et~al.}(2018)\citenamefont {Tomita}, \citenamefont {Kurosawa}, \citenamefont {Ueda},\ and\ \citenamefont {Sawada}}]{tomita2018metamaterials}%
  \BibitemOpen
  \bibfield  {author} {\bibinfo {author} {\bibfnamefont {S.}~\bibnamefont {Tomita}}, \bibinfo {author} {\bibfnamefont {H.}~\bibnamefont {Kurosawa}}, \bibinfo {author} {\bibfnamefont {T.}~\bibnamefont {Ueda}}, \ and\ \bibinfo {author} {\bibfnamefont {K.}~\bibnamefont {Sawada}},\ }\bibfield  {title} {\enquote {\bibinfo {title} {Metamaterials with magnetism and chirality},}\ }\href@noop {} {\bibfield  {journal} {\bibinfo  {journal} {Journal of Physics D: Applied Physics}\ }\textbf {\bibinfo {volume} {51}},\ \bibinfo {pages} {083001} (\bibinfo {year} {2018})}\BibitemShut {NoStop}%
\bibitem [{\citenamefont {Thiaville}\ and\ \citenamefont {Fert}(1992)}]{thiaville1992twisted}%
  \BibitemOpen
  \bibfield  {author} {\bibinfo {author} {\bibfnamefont {A.}~\bibnamefont {Thiaville}}\ and\ \bibinfo {author} {\bibfnamefont {A.}~\bibnamefont {Fert}},\ }\bibfield  {title} {\enquote {\bibinfo {title} {Twisted spin configurations in thin magnetic layers with interface anisotropy},}\ }\href@noop {} {\bibfield  {journal} {\bibinfo  {journal} {Journal of magnetism and magnetic materials}\ }\textbf {\bibinfo {volume} {113}},\ \bibinfo {pages} {161--172} (\bibinfo {year} {1992})}\BibitemShut {NoStop}%
\bibitem [{\citenamefont {Dzyaloshinskii}(1960)}]{dzyaloshinskii1960}%
  \BibitemOpen
  \bibfield  {author} {\bibinfo {author} {\bibfnamefont {I.~E.}\ \bibnamefont {Dzyaloshinskii}},\ }\bibfield  {title} {\enquote {\bibinfo {title} {On the magneto-electrical effects in antiferromagnets},}\ }\href@noop {} {\bibfield  {journal} {\bibinfo  {journal} {Soviet Physics JETP}\ }\textbf {\bibinfo {volume} {10}},\ \bibinfo {pages} {628--629} (\bibinfo {year} {1960})}\BibitemShut {NoStop}%
\bibitem [{\citenamefont {Dzyaloshinskii}(1958)}]{dzyaloshinskii1958}%
  \BibitemOpen
  \bibfield  {author} {\bibinfo {author} {\bibfnamefont {I.}~\bibnamefont {Dzyaloshinskii}},\ }\bibfield  {title} {\enquote {\bibinfo {title} {A thermodynamic theory of weak ferromagnetism of antiferromagnetics},}\ }\href@noop {} {\bibfield  {journal} {\bibinfo  {journal} {J. Phys. Chem. Solids}\ }\textbf {\bibinfo {volume} {4}},\ \bibinfo {pages} {241} (\bibinfo {year} {1958})}\BibitemShut {NoStop}%
\bibitem [{\citenamefont {Togawa}\ \emph {et~al.}(2016)\citenamefont {Togawa}, \citenamefont {Kousaka}, \citenamefont {Inoue},\ and\ \citenamefont {Kishine}}]{togawa2016symmetry}%
  \BibitemOpen
  \bibfield  {author} {\bibinfo {author} {\bibfnamefont {Y.}~\bibnamefont {Togawa}}, \bibinfo {author} {\bibfnamefont {Y.}~\bibnamefont {Kousaka}}, \bibinfo {author} {\bibfnamefont {K.}~\bibnamefont {Inoue}}, \ and\ \bibinfo {author} {\bibfnamefont {J.-i.}\ \bibnamefont {Kishine}},\ }\bibfield  {title} {\enquote {\bibinfo {title} {Symmetry, structure, and dynamics of monoaxial chiral magnets},}\ }\href@noop {} {\bibfield  {journal} {\bibinfo  {journal} {Journal of the Physical Society of Japan}\ }\textbf {\bibinfo {volume} {85}},\ \bibinfo {pages} {112001} (\bibinfo {year} {2016})}\BibitemShut {NoStop}%
\bibitem [{\citenamefont {Lucassen}\ \emph {et~al.}(2019)\citenamefont {Lucassen}, \citenamefont {Meijer}, \citenamefont {Kurnosikov}, \citenamefont {Swagten}, \citenamefont {Koopmans}, \citenamefont {Lavrijsen}, \citenamefont {Kloodt-Twesten}, \citenamefont {Fr{\"o}mter},\ and\ \citenamefont {Duine}}]{lucassen2019tuning}%
  \BibitemOpen
  \bibfield  {author} {\bibinfo {author} {\bibfnamefont {J.}~\bibnamefont {Lucassen}}, \bibinfo {author} {\bibfnamefont {M.~J.}\ \bibnamefont {Meijer}}, \bibinfo {author} {\bibfnamefont {O.}~\bibnamefont {Kurnosikov}}, \bibinfo {author} {\bibfnamefont {H.~J.}\ \bibnamefont {Swagten}}, \bibinfo {author} {\bibfnamefont {B.}~\bibnamefont {Koopmans}}, \bibinfo {author} {\bibfnamefont {R.}~\bibnamefont {Lavrijsen}}, \bibinfo {author} {\bibfnamefont {F.}~\bibnamefont {Kloodt-Twesten}}, \bibinfo {author} {\bibfnamefont {R.}~\bibnamefont {Fr{\"o}mter}}, \ and\ \bibinfo {author} {\bibfnamefont {R.~A.}\ \bibnamefont {Duine}},\ }\bibfield  {title} {\enquote {\bibinfo {title} {Tuning magnetic chirality by dipolar interactions},}\ }\href@noop {} {\bibfield  {journal} {\bibinfo  {journal} {Physical review letters}\ }\textbf {\bibinfo {volume} {123}},\ \bibinfo {pages} {157201} (\bibinfo {year} {2019})}\BibitemShut {NoStop}%
\bibitem [{\citenamefont {Mellado}\ \emph {et~al.}(2023)\citenamefont {Mellado}, \citenamefont {Concha}, \citenamefont {Hofhuis},\ and\ \citenamefont {Tapia}}]{mellado2023intrinsic}%
  \BibitemOpen
  \bibfield  {author} {\bibinfo {author} {\bibfnamefont {P.}~\bibnamefont {Mellado}}, \bibinfo {author} {\bibfnamefont {A.}~\bibnamefont {Concha}}, \bibinfo {author} {\bibfnamefont {K.}~\bibnamefont {Hofhuis}}, \ and\ \bibinfo {author} {\bibfnamefont {I.}~\bibnamefont {Tapia}},\ }\bibfield  {title} {\enquote {\bibinfo {title} {Intrinsic chiral field as vector potential of the magnetic current in the zig-zag lattice of magnetic dipoles},}\ }\href@noop {} {\bibfield  {journal} {\bibinfo  {journal} {Scientific reports}\ }\textbf {\bibinfo {volume} {13}},\ \bibinfo {pages} {1245} (\bibinfo {year} {2023})}\BibitemShut {NoStop}%
\bibitem [{\citenamefont {Paula}\ and\ \citenamefont {Tapia}(2023)}]{paula2023magnetic}%
  \BibitemOpen
  \bibfield  {author} {\bibinfo {author} {\bibfnamefont {M.}~\bibnamefont {Paula}}\ and\ \bibinfo {author} {\bibfnamefont {I.}~\bibnamefont {Tapia}},\ }\bibfield  {title} {\enquote {\bibinfo {title} {Magnetic solitons due to interfacial chiral interactions},}\ }\href@noop {} {\bibfield  {journal} {\bibinfo  {journal} {Journal of Physics: Condensed Matter}\ }\textbf {\bibinfo {volume} {35}},\ \bibinfo {pages} {164002} (\bibinfo {year} {2023})}\BibitemShut {NoStop}%
\bibitem [{\citenamefont {Yu}\ and\ \citenamefont {Bauer}(2021)}]{yu2021chiral}%
  \BibitemOpen
  \bibfield  {author} {\bibinfo {author} {\bibfnamefont {T.}~\bibnamefont {Yu}}\ and\ \bibinfo {author} {\bibfnamefont {G.~E.}\ \bibnamefont {Bauer}},\ }\bibfield  {title} {\enquote {\bibinfo {title} {Chiral coupling to magnetodipolar radiation},}\ }\href@noop {} {\bibfield  {journal} {\bibinfo  {journal} {Chirality, Magnetism and Magnetoelectricity: Separate Phenomena and Joint Effects in Metamaterial Structures}\ ,\ \bibinfo {pages} {1--23}} (\bibinfo {year} {2021})}\BibitemShut {NoStop}%
\bibitem [{\citenamefont {Ray}\ and\ \citenamefont {Das}(2021)}]{ray2021hierarchy}%
  \BibitemOpen
  \bibfield  {author} {\bibinfo {author} {\bibfnamefont {S.}~\bibnamefont {Ray}}\ and\ \bibinfo {author} {\bibfnamefont {T.}~\bibnamefont {Das}},\ }\bibfield  {title} {\enquote {\bibinfo {title} {Hierarchy of multi-order skyrmion phases in twisted magnetic bilayers},}\ }\href@noop {} {\bibfield  {journal} {\bibinfo  {journal} {Physical Review B}\ }\textbf {\bibinfo {volume} {104}},\ \bibinfo {pages} {014410} (\bibinfo {year} {2021})}\BibitemShut {NoStop}%
\bibitem [{\citenamefont {Shindou}\ \emph {et~al.}(2013)\citenamefont {Shindou}, \citenamefont {Ohe}, \citenamefont {Matsumoto}, \citenamefont {Murakami},\ and\ \citenamefont {Saitoh}}]{shindou2013chiral}%
  \BibitemOpen
  \bibfield  {author} {\bibinfo {author} {\bibfnamefont {R.}~\bibnamefont {Shindou}}, \bibinfo {author} {\bibfnamefont {J.-i.}\ \bibnamefont {Ohe}}, \bibinfo {author} {\bibfnamefont {R.}~\bibnamefont {Matsumoto}}, \bibinfo {author} {\bibfnamefont {S.}~\bibnamefont {Murakami}}, \ and\ \bibinfo {author} {\bibfnamefont {E.}~\bibnamefont {Saitoh}},\ }\bibfield  {title} {\enquote {\bibinfo {title} {Chiral spin-wave edge modes in dipolar magnetic thin films},}\ }\href@noop {} {\bibfield  {journal} {\bibinfo  {journal} {Physical Review B}\ }\textbf {\bibinfo {volume} {87}},\ \bibinfo {pages} {174402} (\bibinfo {year} {2013})}\BibitemShut {NoStop}%
\bibitem [{\citenamefont {Malozemoff}\ and\ \citenamefont {Slonczewski}(2013)}]{malozemoff2013magnetic}%
  \BibitemOpen
  \bibfield  {author} {\bibinfo {author} {\bibfnamefont {A.~P.}\ \bibnamefont {Malozemoff}}\ and\ \bibinfo {author} {\bibfnamefont {J.~C.}\ \bibnamefont {Slonczewski}},\ }\href@noop {} {\emph {\bibinfo {title} {Magnetic domain walls in bubble materials: advances in materials and device research}}},\ Vol.~\bibinfo {volume} {1}\ (\bibinfo  {publisher} {Academic press},\ \bibinfo {year} {2013})\BibitemShut {NoStop}%
\bibitem [{\citenamefont {Hubert}\ and\ \citenamefont {Sch{\"a}fer}(2008)}]{hubert2008magnetic}%
  \BibitemOpen
  \bibfield  {author} {\bibinfo {author} {\bibfnamefont {A.}~\bibnamefont {Hubert}}\ and\ \bibinfo {author} {\bibfnamefont {R.}~\bibnamefont {Sch{\"a}fer}},\ }\href@noop {} {\emph {\bibinfo {title} {Magnetic domains: the analysis of magnetic microstructures}}}\ (\bibinfo  {publisher} {Springer Science \& Business Media},\ \bibinfo {year} {2008})\BibitemShut {NoStop}%
\bibitem [{\citenamefont {Tapia}, \citenamefont {Cazor},\ and\ \citenamefont {Mellado}(2024)}]{tapia2024chiral}%
  \BibitemOpen
  \bibfield  {author} {\bibinfo {author} {\bibfnamefont {I.}~\bibnamefont {Tapia}}, \bibinfo {author} {\bibfnamefont {X.}~\bibnamefont {Cazor}}, \ and\ \bibinfo {author} {\bibfnamefont {P.}~\bibnamefont {Mellado}},\ }\bibfield  {title} {\enquote {\bibinfo {title} {Chiral magnetic phases in moire bilayers of magnetic dipoles},}\ }\href@noop {} {\bibfield  {journal} {\bibinfo  {journal} {Advanced Physics Research}\ }\textbf {\bibinfo {volume} {3}},\ \bibinfo {pages} {2300135} (\bibinfo {year} {2024})}\BibitemShut {NoStop}%
\bibitem [{\citenamefont {Tokura}\ and\ \citenamefont {Nagaosa}(2018)}]{tokura2018nonreciprocal}%
  \BibitemOpen
  \bibfield  {author} {\bibinfo {author} {\bibfnamefont {Y.}~\bibnamefont {Tokura}}\ and\ \bibinfo {author} {\bibfnamefont {N.}~\bibnamefont {Nagaosa}},\ }\bibfield  {title} {\enquote {\bibinfo {title} {Nonreciprocal responses from non-centrosymmetric quantum materials},}\ }\href@noop {} {\bibfield  {journal} {\bibinfo  {journal} {Nature communications}\ }\textbf {\bibinfo {volume} {9}},\ \bibinfo {pages} {3740} (\bibinfo {year} {2018})}\BibitemShut {NoStop}%
\bibitem [{\citenamefont {Kawamura}(2010)}]{kawamura2010chirality}%
  \BibitemOpen
  \bibfield  {author} {\bibinfo {author} {\bibfnamefont {H.}~\bibnamefont {Kawamura}},\ }\bibfield  {title} {\enquote {\bibinfo {title} {Chirality scenario of the spin-glass ordering},}\ }\href@noop {} {\bibfield  {journal} {\bibinfo  {journal} {Journal of the Physical Society of Japan}\ }\textbf {\bibinfo {volume} {79}},\ \bibinfo {pages} {011007} (\bibinfo {year} {2010})}\BibitemShut {NoStop}%
\bibitem [{\citenamefont {Gong}\ \emph {et~al.}(2017)\citenamefont {Gong}, \citenamefont {Li}, \citenamefont {Li}, \citenamefont {Ji}, \citenamefont {Stern}, \citenamefont {Xia}, \citenamefont {Cao}, \citenamefont {Bao}, \citenamefont {Wang}, \citenamefont {Wang} \emph {et~al.}}]{gong2017discovery}%
  \BibitemOpen
  \bibfield  {author} {\bibinfo {author} {\bibfnamefont {C.}~\bibnamefont {Gong}}, \bibinfo {author} {\bibfnamefont {L.}~\bibnamefont {Li}}, \bibinfo {author} {\bibfnamefont {Z.}~\bibnamefont {Li}}, \bibinfo {author} {\bibfnamefont {H.}~\bibnamefont {Ji}}, \bibinfo {author} {\bibfnamefont {A.}~\bibnamefont {Stern}}, \bibinfo {author} {\bibfnamefont {Y.}~\bibnamefont {Xia}}, \bibinfo {author} {\bibfnamefont {T.}~\bibnamefont {Cao}}, \bibinfo {author} {\bibfnamefont {W.}~\bibnamefont {Bao}}, \bibinfo {author} {\bibfnamefont {C.}~\bibnamefont {Wang}}, \bibinfo {author} {\bibfnamefont {Y.}~\bibnamefont {Wang}},  \emph {et~al.},\ }\bibfield  {title} {\enquote {\bibinfo {title} {Discovery of intrinsic ferromagnetism in two-dimensional van der waals crystals},}\ }\href@noop {} {\bibfield  {journal} {\bibinfo  {journal} {Nature}\ }\textbf {\bibinfo {volume} {546}},\ \bibinfo {pages} {265--269} (\bibinfo {year} {2017})}\BibitemShut {NoStop}%
\bibitem [{\citenamefont {Chen}\ \emph {et~al.}(2013)\citenamefont {Chen}, \citenamefont {Ma}, \citenamefont {N’Diaye}, \citenamefont {Kwon}, \citenamefont {Won}, \citenamefont {Wu},\ and\ \citenamefont {Schmid}}]{chen2013tailoring}%
  \BibitemOpen
  \bibfield  {author} {\bibinfo {author} {\bibfnamefont {G.}~\bibnamefont {Chen}}, \bibinfo {author} {\bibfnamefont {T.}~\bibnamefont {Ma}}, \bibinfo {author} {\bibfnamefont {A.~T.}\ \bibnamefont {N’Diaye}}, \bibinfo {author} {\bibfnamefont {H.}~\bibnamefont {Kwon}}, \bibinfo {author} {\bibfnamefont {C.}~\bibnamefont {Won}}, \bibinfo {author} {\bibfnamefont {Y.}~\bibnamefont {Wu}}, \ and\ \bibinfo {author} {\bibfnamefont {A.~K.}\ \bibnamefont {Schmid}},\ }\bibfield  {title} {\enquote {\bibinfo {title} {Tailoring the chirality of magnetic domain walls by interface engineering},}\ }\href@noop {} {\bibfield  {journal} {\bibinfo  {journal} {Nature communications}\ }\textbf {\bibinfo {volume} {4}},\ \bibinfo {pages} {2671} (\bibinfo {year} {2013})}\BibitemShut {NoStop}%
\bibitem [{\citenamefont {Elitzur}, \citenamefont {Pearson},\ and\ \citenamefont {Shigemitsu}(1979)}]{elitzur1979phase}%
  \BibitemOpen
  \bibfield  {author} {\bibinfo {author} {\bibfnamefont {S.}~\bibnamefont {Elitzur}}, \bibinfo {author} {\bibfnamefont {R.}~\bibnamefont {Pearson}}, \ and\ \bibinfo {author} {\bibfnamefont {J.}~\bibnamefont {Shigemitsu}},\ }\bibfield  {title} {\enquote {\bibinfo {title} {Phase structure of discrete abelian spin and gauge systems},}\ }\href@noop {} {\bibfield  {journal} {\bibinfo  {journal} {Physical Review D}\ }\textbf {\bibinfo {volume} {19}},\ \bibinfo {pages} {3698} (\bibinfo {year} {1979})}\BibitemShut {NoStop}%
\bibitem [{\citenamefont {Sun}\ \emph {et~al.}(2019)\citenamefont {Sun}, \citenamefont {Vekua}, \citenamefont {Cobanera},\ and\ \citenamefont {Ortiz}}]{sun2019phase}%
  \BibitemOpen
  \bibfield  {author} {\bibinfo {author} {\bibfnamefont {G.}~\bibnamefont {Sun}}, \bibinfo {author} {\bibfnamefont {T.}~\bibnamefont {Vekua}}, \bibinfo {author} {\bibfnamefont {E.}~\bibnamefont {Cobanera}}, \ and\ \bibinfo {author} {\bibfnamefont {G.}~\bibnamefont {Ortiz}},\ }\bibfield  {title} {\enquote {\bibinfo {title} {Phase transitions in the z p and u (1) clock models},}\ }\href@noop {} {\bibfield  {journal} {\bibinfo  {journal} {Physical Review B}\ }\textbf {\bibinfo {volume} {100}},\ \bibinfo {pages} {094428} (\bibinfo {year} {2019})}\BibitemShut {NoStop}%
\bibitem [{\citenamefont {Jos{\'e}}\ \emph {et~al.}(1977)\citenamefont {Jos{\'e}}, \citenamefont {Kadanoff}, \citenamefont {Kirkpatrick},\ and\ \citenamefont {Nelson}}]{jose1977renormalization}%
  \BibitemOpen
  \bibfield  {author} {\bibinfo {author} {\bibfnamefont {J.~V.}\ \bibnamefont {Jos{\'e}}}, \bibinfo {author} {\bibfnamefont {L.~P.}\ \bibnamefont {Kadanoff}}, \bibinfo {author} {\bibfnamefont {S.}~\bibnamefont {Kirkpatrick}}, \ and\ \bibinfo {author} {\bibfnamefont {D.~R.}\ \bibnamefont {Nelson}},\ }\bibfield  {title} {\enquote {\bibinfo {title} {Renormalization, vortices, and symmetry-breaking perturbations in the two-dimensional planar model},}\ }\href@noop {} {\bibfield  {journal} {\bibinfo  {journal} {Physical Review B}\ }\textbf {\bibinfo {volume} {16}},\ \bibinfo {pages} {1217} (\bibinfo {year} {1977})}\BibitemShut {NoStop}%
\bibitem [{\citenamefont {Villain}(1975)}]{villain1975theory}%
  \BibitemOpen
  \bibfield  {author} {\bibinfo {author} {\bibfnamefont {J.}~\bibnamefont {Villain}},\ }\bibfield  {title} {\enquote {\bibinfo {title} {Theory of one-and two-dimensional magnets with an easy magnetization plane. ii. the planar, classical, two-dimensional magnet},}\ }\href@noop {} {\bibfield  {journal} {\bibinfo  {journal} {Journal de Physique}\ }\textbf {\bibinfo {volume} {36}},\ \bibinfo {pages} {581--590} (\bibinfo {year} {1975})}\BibitemShut {NoStop}%
\bibitem [{\citenamefont {Cardy}(1996)}]{cardy1996scaling}%
  \BibitemOpen
  \bibfield  {author} {\bibinfo {author} {\bibfnamefont {J.}~\bibnamefont {Cardy}},\ }\href@noop {} {\emph {\bibinfo {title} {Scaling and renormalization in statistical physics}}},\ Vol.~\bibinfo {volume} {5}\ (\bibinfo  {publisher} {Cambridge university press},\ \bibinfo {year} {1996})\BibitemShut {NoStop}%
\bibitem [{\citenamefont {Baek}\ \emph {et~al.}(2013)\citenamefont {Baek}, \citenamefont {M{\"a}kel{\"a}}, \citenamefont {Minnhagen},\ and\ \citenamefont {Kim}}]{baek2013residual}%
  \BibitemOpen
  \bibfield  {author} {\bibinfo {author} {\bibfnamefont {S.~K.}\ \bibnamefont {Baek}}, \bibinfo {author} {\bibfnamefont {H.}~\bibnamefont {M{\"a}kel{\"a}}}, \bibinfo {author} {\bibfnamefont {P.}~\bibnamefont {Minnhagen}}, \ and\ \bibinfo {author} {\bibfnamefont {B.~J.}\ \bibnamefont {Kim}},\ }\bibfield  {title} {\enquote {\bibinfo {title} {Residual discrete symmetry of the five-state clock model},}\ }\href@noop {} {\bibfield  {journal} {\bibinfo  {journal} {Physical review. E, Statistical, nonlinear, and soft matter physics}\ }\textbf {\bibinfo {volume} {88}},\ \bibinfo {pages} {012125} (\bibinfo {year} {2013})}\BibitemShut {NoStop}%
\bibitem [{\citenamefont {Nussinov}\ and\ \citenamefont {Van Den~Brink}(2015)}]{nussinov2015compass}%
  \BibitemOpen
  \bibfield  {author} {\bibinfo {author} {\bibfnamefont {Z.}~\bibnamefont {Nussinov}}\ and\ \bibinfo {author} {\bibfnamefont {J.}~\bibnamefont {Van Den~Brink}},\ }\bibfield  {title} {\enquote {\bibinfo {title} {Compass models: Theory and physical motivations},}\ }\href@noop {} {\bibfield  {journal} {\bibinfo  {journal} {Reviews of Modern Physics}\ }\textbf {\bibinfo {volume} {87}},\ \bibinfo {pages} {1--59} (\bibinfo {year} {2015})}\BibitemShut {NoStop}%
\bibitem [{\citenamefont {Kim}(2017)}]{kim2017partition}%
  \BibitemOpen
  \bibfield  {author} {\bibinfo {author} {\bibfnamefont {D.-H.}\ \bibnamefont {Kim}},\ }\bibfield  {title} {\enquote {\bibinfo {title} {Partition function zeros of the p-state clock model in the complex temperature plane},}\ }\href@noop {} {\bibfield  {journal} {\bibinfo  {journal} {Physical Review E}\ }\textbf {\bibinfo {volume} {96}},\ \bibinfo {pages} {052130} (\bibinfo {year} {2017})}\BibitemShut {NoStop}%
\bibitem [{\citenamefont {Gopinathan}\ and\ \citenamefont {Grier}(2004)}]{gopinathan2004statistically}%
  \BibitemOpen
  \bibfield  {author} {\bibinfo {author} {\bibfnamefont {A.}~\bibnamefont {Gopinathan}}\ and\ \bibinfo {author} {\bibfnamefont {D.~G.}\ \bibnamefont {Grier}},\ }\bibfield  {title} {\enquote {\bibinfo {title} {Statistically locked-in transport through periodic potential landscapes},}\ }\href@noop {} {\bibfield  {journal} {\bibinfo  {journal} {Physical review letters}\ }\textbf {\bibinfo {volume} {92}},\ \bibinfo {pages} {130602} (\bibinfo {year} {2004})}\BibitemShut {NoStop}%
\bibitem [{\citenamefont {Amit}, \citenamefont {Goldschmidt},\ and\ \citenamefont {Grinstein}(1980)}]{amit1980renormalisation}%
  \BibitemOpen
  \bibfield  {author} {\bibinfo {author} {\bibfnamefont {D.~J.}\ \bibnamefont {Amit}}, \bibinfo {author} {\bibfnamefont {Y.~Y.}\ \bibnamefont {Goldschmidt}}, \ and\ \bibinfo {author} {\bibfnamefont {S.}~\bibnamefont {Grinstein}},\ }\bibfield  {title} {\enquote {\bibinfo {title} {Renormalisation group analysis of the phase transition in the 2d coulomb gas, sine-gordon theory and xy-model},}\ }\href@noop {} {\bibfield  {journal} {\bibinfo  {journal} {Journal of Physics A: Mathematical and General}\ }\textbf {\bibinfo {volume} {13}},\ \bibinfo {pages} {585} (\bibinfo {year} {1980})}\BibitemShut {NoStop}%
\bibitem [{\citenamefont {Lemmens}, \citenamefont {Kimura},\ and\ \citenamefont {De~Jonge}(1986)}]{lemmens1986sine}%
  \BibitemOpen
  \bibfield  {author} {\bibinfo {author} {\bibfnamefont {L.}~\bibnamefont {Lemmens}}, \bibinfo {author} {\bibfnamefont {I.}~\bibnamefont {Kimura}}, \ and\ \bibinfo {author} {\bibfnamefont {W.}~\bibnamefont {De~Jonge}},\ }\bibfield  {title} {\enquote {\bibinfo {title} {Sine-gordon kink solitons and the magnetisation in one-dimensional antiferromagnetic chains},}\ }\href@noop {} {\bibfield  {journal} {\bibinfo  {journal} {Journal of Physics C: Solid State Physics}\ }\textbf {\bibinfo {volume} {19}},\ \bibinfo {pages} {139} (\bibinfo {year} {1986})}\BibitemShut {NoStop}%
\bibitem [{\citenamefont {Concha}, \citenamefont {Aguayo},\ and\ \citenamefont {Mellado}(2018)}]{concha2018designing}%
  \BibitemOpen
  \bibfield  {author} {\bibinfo {author} {\bibfnamefont {A.}~\bibnamefont {Concha}}, \bibinfo {author} {\bibfnamefont {D.}~\bibnamefont {Aguayo}}, \ and\ \bibinfo {author} {\bibfnamefont {P.}~\bibnamefont {Mellado}},\ }\bibfield  {title} {\enquote {\bibinfo {title} {Designing hysteresis with dipolar chains},}\ }\href@noop {} {\bibfield  {journal} {\bibinfo  {journal} {Physical Review Letters}\ }\textbf {\bibinfo {volume} {120}},\ \bibinfo {pages} {157202} (\bibinfo {year} {2018})}\BibitemShut {NoStop}%
\bibitem [{\citenamefont {Cheong}\ and\ \citenamefont {Xu}(2022)}]{cheong2022magnetic}%
  \BibitemOpen
  \bibfield  {author} {\bibinfo {author} {\bibfnamefont {S.-W.}\ \bibnamefont {Cheong}}\ and\ \bibinfo {author} {\bibfnamefont {X.}~\bibnamefont {Xu}},\ }\bibfield  {title} {\enquote {\bibinfo {title} {Magnetic chirality},}\ }\href@noop {} {\bibfield  {journal} {\bibinfo  {journal} {npj Quantum Materials}\ }\textbf {\bibinfo {volume} {7}},\ \bibinfo {pages} {40} (\bibinfo {year} {2022})}\BibitemShut {NoStop}%
\bibitem [{\citenamefont {Batista}\ \emph {et~al.}(2016)\citenamefont {Batista}, \citenamefont {Lin}, \citenamefont {Hayami},\ and\ \citenamefont {Kamiya}}]{batista2016frustration}%
  \BibitemOpen
  \bibfield  {author} {\bibinfo {author} {\bibfnamefont {C.~D.}\ \bibnamefont {Batista}}, \bibinfo {author} {\bibfnamefont {S.-Z.}\ \bibnamefont {Lin}}, \bibinfo {author} {\bibfnamefont {S.}~\bibnamefont {Hayami}}, \ and\ \bibinfo {author} {\bibfnamefont {Y.}~\bibnamefont {Kamiya}},\ }\bibfield  {title} {\enquote {\bibinfo {title} {Frustration and chiral orderings in correlated electron systems},}\ }\href@noop {} {\bibfield  {journal} {\bibinfo  {journal} {Reports on Progress in Physics}\ }\textbf {\bibinfo {volume} {79}},\ \bibinfo {pages} {084504} (\bibinfo {year} {2016})}\BibitemShut {NoStop}%
\bibitem [{\citenamefont {Yambe}\ and\ \citenamefont {Hayami}(2023)}]{yambe2023ferrochiral}%
  \BibitemOpen
  \bibfield  {author} {\bibinfo {author} {\bibfnamefont {R.}~\bibnamefont {Yambe}}\ and\ \bibinfo {author} {\bibfnamefont {S.}~\bibnamefont {Hayami}},\ }\bibfield  {title} {\enquote {\bibinfo {title} {Ferrochiral, antiferrochiral, and ferrichiral skyrmion crystals in an itinerant honeycomb magnet},}\ }\href@noop {} {\bibfield  {journal} {\bibinfo  {journal} {Physical Review B}\ }\textbf {\bibinfo {volume} {107}},\ \bibinfo {pages} {014417} (\bibinfo {year} {2023})}\BibitemShut {NoStop}%
\bibitem [{\citenamefont {Whitsitt}, \citenamefont {Samajdar},\ and\ \citenamefont {Sachdev}(2018)}]{whitsitt2018quantum}%
  \BibitemOpen
  \bibfield  {author} {\bibinfo {author} {\bibfnamefont {S.}~\bibnamefont {Whitsitt}}, \bibinfo {author} {\bibfnamefont {R.}~\bibnamefont {Samajdar}}, \ and\ \bibinfo {author} {\bibfnamefont {S.}~\bibnamefont {Sachdev}},\ }\bibfield  {title} {\enquote {\bibinfo {title} {Quantum field theory for the chiral clock transition in one spatial dimension},}\ }\href@noop {} {\bibfield  {journal} {\bibinfo  {journal} {Physical Review B}\ }\textbf {\bibinfo {volume} {98}},\ \bibinfo {pages} {205118} (\bibinfo {year} {2018})}\BibitemShut {NoStop}%
\bibitem [{\citenamefont {Maryasin}, \citenamefont {Zhitomirsky},\ and\ \citenamefont {Moessner}(2016)}]{maryasin2016low}%
  \BibitemOpen
  \bibfield  {author} {\bibinfo {author} {\bibfnamefont {V.}~\bibnamefont {Maryasin}}, \bibinfo {author} {\bibfnamefont {M.~E.}\ \bibnamefont {Zhitomirsky}}, \ and\ \bibinfo {author} {\bibfnamefont {R.}~\bibnamefont {Moessner}},\ }\bibfield  {title} {\enquote {\bibinfo {title} {Low-field behavior of an xy pyrochlore antiferromagnet: Emergent clock anisotropies},}\ }\href@noop {} {\bibfield  {journal} {\bibinfo  {journal} {Physical Review B}\ }\textbf {\bibinfo {volume} {93}},\ \bibinfo {pages} {100406} (\bibinfo {year} {2016})}\BibitemShut {NoStop}%
\bibitem [{\citenamefont {Hejazi}, \citenamefont {Luo},\ and\ \citenamefont {Balents}(2020)}]{hejazi2020noncollinear}%
  \BibitemOpen
  \bibfield  {author} {\bibinfo {author} {\bibfnamefont {K.}~\bibnamefont {Hejazi}}, \bibinfo {author} {\bibfnamefont {Z.-X.}\ \bibnamefont {Luo}}, \ and\ \bibinfo {author} {\bibfnamefont {L.}~\bibnamefont {Balents}},\ }\bibfield  {title} {\enquote {\bibinfo {title} {Noncollinear phases in moir{\'e} magnets},}\ }\href@noop {} {\bibfield  {journal} {\bibinfo  {journal} {Proceedings of the National Academy of Sciences}\ }\textbf {\bibinfo {volume} {117}},\ \bibinfo {pages} {10721--10726} (\bibinfo {year} {2020})}\BibitemShut {NoStop}%
\bibitem [{\citenamefont {Vignaud}\ \emph {et~al.}(2023)\citenamefont {Vignaud}, \citenamefont {Perconte}, \citenamefont {Yang}, \citenamefont {Kousar}, \citenamefont {Wagner}, \citenamefont {Gay}, \citenamefont {Watanabe}, \citenamefont {Taniguchi}, \citenamefont {Courtois}, \citenamefont {Han} \emph {et~al.}}]{vignaud2023evidence}%
  \BibitemOpen
  \bibfield  {author} {\bibinfo {author} {\bibfnamefont {H.}~\bibnamefont {Vignaud}}, \bibinfo {author} {\bibfnamefont {D.}~\bibnamefont {Perconte}}, \bibinfo {author} {\bibfnamefont {W.}~\bibnamefont {Yang}}, \bibinfo {author} {\bibfnamefont {B.}~\bibnamefont {Kousar}}, \bibinfo {author} {\bibfnamefont {E.}~\bibnamefont {Wagner}}, \bibinfo {author} {\bibfnamefont {F.}~\bibnamefont {Gay}}, \bibinfo {author} {\bibfnamefont {K.}~\bibnamefont {Watanabe}}, \bibinfo {author} {\bibfnamefont {T.}~\bibnamefont {Taniguchi}}, \bibinfo {author} {\bibfnamefont {H.}~\bibnamefont {Courtois}}, \bibinfo {author} {\bibfnamefont {Z.}~\bibnamefont {Han}},  \emph {et~al.},\ }\bibfield  {title} {\enquote {\bibinfo {title} {Evidence for chiral supercurrent in quantum hall josephson junctions},}\ }\href@noop {} {\bibfield  {journal} {\bibinfo  {journal} {Nature}\ }\textbf {\bibinfo {volume} {624}},\ \bibinfo {pages} {545--550} (\bibinfo {year} {2023})}\BibitemShut {NoStop}%
\bibitem [{\citenamefont {Cuevas-Maraver}, \citenamefont {Kevrekidis},\ and\ \citenamefont {Williams}(2014)}]{cuevas2014sine}%
  \BibitemOpen
  \bibfield  {author} {\bibinfo {author} {\bibfnamefont {J.}~\bibnamefont {Cuevas-Maraver}}, \bibinfo {author} {\bibfnamefont {P.~G.}\ \bibnamefont {Kevrekidis}}, \ and\ \bibinfo {author} {\bibfnamefont {F.}~\bibnamefont {Williams}},\ }\bibfield  {title} {\enquote {\bibinfo {title} {The sine-gordon model and its applications},}\ }\href@noop {} {\bibfield  {journal} {\bibinfo  {journal} {Nonlinear systems and complexity}\ }\textbf {\bibinfo {volume} {10}} (\bibinfo {year} {2014})}\BibitemShut {NoStop}%
\bibitem [{\citenamefont {Flamino}\ and\ \citenamefont {Giedt}(2020)}]{flamino2020lattice}%
  \BibitemOpen
  \bibfield  {author} {\bibinfo {author} {\bibfnamefont {J.}~\bibnamefont {Flamino}}\ and\ \bibinfo {author} {\bibfnamefont {J.}~\bibnamefont {Giedt}},\ }\bibfield  {title} {\enquote {\bibinfo {title} {Lattice sine-gordon model},}\ }\href@noop {} {\bibfield  {journal} {\bibinfo  {journal} {Physical Review D}\ }\textbf {\bibinfo {volume} {101}},\ \bibinfo {pages} {074503} (\bibinfo {year} {2020})}\BibitemShut {NoStop}%
\bibitem [{\citenamefont {Coppersmith}\ and\ \citenamefont {Fisher}(1983)}]{coppersmith1983pinning}%
  \BibitemOpen
  \bibfield  {author} {\bibinfo {author} {\bibfnamefont {S.}~\bibnamefont {Coppersmith}}\ and\ \bibinfo {author} {\bibfnamefont {D.}~\bibnamefont {Fisher}},\ }\bibfield  {title} {\enquote {\bibinfo {title} {Pinning transition of the discrete sine-gordon equation},}\ }\href@noop {} {\bibfield  {journal} {\bibinfo  {journal} {Physical Review B}\ }\textbf {\bibinfo {volume} {28}},\ \bibinfo {pages} {2566} (\bibinfo {year} {1983})}\BibitemShut {NoStop}%
\bibitem [{\citenamefont {Miller}(1986)}]{miller1986nonlinear}%
  \BibitemOpen
  \bibfield  {author} {\bibinfo {author} {\bibfnamefont {M.}~\bibnamefont {Miller}},\ }\bibfield  {title} {\enquote {\bibinfo {title} {Nonlinear wave propagation in periodic systems: The driven sine-gordon chain},}\ }\href@noop {} {\bibfield  {journal} {\bibinfo  {journal} {Physical Review B}\ }\textbf {\bibinfo {volume} {33}},\ \bibinfo {pages} {1641} (\bibinfo {year} {1986})}\BibitemShut {NoStop}%
\bibitem [{\citenamefont {Bruce}, \citenamefont {Cowley},\ and\ \citenamefont {Murray}(1978)}]{bruce1978theory}%
  \BibitemOpen
  \bibfield  {author} {\bibinfo {author} {\bibfnamefont {A.}~\bibnamefont {Bruce}}, \bibinfo {author} {\bibfnamefont {R.}~\bibnamefont {Cowley}}, \ and\ \bibinfo {author} {\bibfnamefont {A.}~\bibnamefont {Murray}},\ }\bibfield  {title} {\enquote {\bibinfo {title} {The theory of structurally incommensurate systems. ii. commensurate-incommensurate phase transitions},}\ }\href@noop {} {\bibfield  {journal} {\bibinfo  {journal} {Journal of Physics C: Solid State Physics}\ }\textbf {\bibinfo {volume} {11}},\ \bibinfo {pages} {3591} (\bibinfo {year} {1978})}\BibitemShut {NoStop}%
\bibitem [{\citenamefont {Popov}\ \emph {et~al.}(2011)\citenamefont {Popov}, \citenamefont {Lebedeva}, \citenamefont {Knizhnik}, \citenamefont {Lozovik},\ and\ \citenamefont {Potapkin}}]{popov2011commensurate}%
  \BibitemOpen
  \bibfield  {author} {\bibinfo {author} {\bibfnamefont {A.~M.}\ \bibnamefont {Popov}}, \bibinfo {author} {\bibfnamefont {I.~V.}\ \bibnamefont {Lebedeva}}, \bibinfo {author} {\bibfnamefont {A.~A.}\ \bibnamefont {Knizhnik}}, \bibinfo {author} {\bibfnamefont {Y.~E.}\ \bibnamefont {Lozovik}}, \ and\ \bibinfo {author} {\bibfnamefont {B.~V.}\ \bibnamefont {Potapkin}},\ }\bibfield  {title} {\enquote {\bibinfo {title} {Commensurate-incommensurate phase transition in bilayer graphene},}\ }\href@noop {} {\bibfield  {journal} {\bibinfo  {journal} {Physical Review B—Condensed Matter and Materials Physics}\ }\textbf {\bibinfo {volume} {84}},\ \bibinfo {pages} {045404} (\bibinfo {year} {2011})}\BibitemShut {NoStop}%
\bibitem [{\citenamefont {Lebedeva}\ and\ \citenamefont {Popov}(2019{\natexlab{a}})}]{lebedeva2019commensurate}%
  \BibitemOpen
  \bibfield  {author} {\bibinfo {author} {\bibfnamefont {I.~V.}\ \bibnamefont {Lebedeva}}\ and\ \bibinfo {author} {\bibfnamefont {A.~M.}\ \bibnamefont {Popov}},\ }\bibfield  {title} {\enquote {\bibinfo {title} {Commensurate-incommensurate phase transition and a network of domain walls in bilayer graphene with a biaxially stretched layer},}\ }\href@noop {} {\bibfield  {journal} {\bibinfo  {journal} {Physical Review B}\ }\textbf {\bibinfo {volume} {99}},\ \bibinfo {pages} {195448} (\bibinfo {year} {2019}{\natexlab{a}})}\BibitemShut {NoStop}%
\bibitem [{\citenamefont {Sasaki}\ and\ \citenamefont {Floria}(1989)}]{sasaki1989symmetry}%
  \BibitemOpen
  \bibfield  {author} {\bibinfo {author} {\bibfnamefont {K.}~\bibnamefont {Sasaki}}\ and\ \bibinfo {author} {\bibfnamefont {L.}~\bibnamefont {Floria}},\ }\bibfield  {title} {\enquote {\bibinfo {title} {Symmetry-breaking commensurate states in generalised frenkel-kontorova models},}\ }\href@noop {} {\bibfield  {journal} {\bibinfo  {journal} {Journal of Physics: Condensed Matter}\ }\textbf {\bibinfo {volume} {1}},\ \bibinfo {pages} {2179} (\bibinfo {year} {1989})}\BibitemShut {NoStop}%
\bibitem [{\citenamefont {Lebedeva}\ and\ \citenamefont {Popov}(2019{\natexlab{b}})}]{lebedeva2019energetics}%
  \BibitemOpen
  \bibfield  {author} {\bibinfo {author} {\bibfnamefont {I.~V.}\ \bibnamefont {Lebedeva}}\ and\ \bibinfo {author} {\bibfnamefont {A.~M.}\ \bibnamefont {Popov}},\ }\bibfield  {title} {\enquote {\bibinfo {title} {Energetics and structure of domain wall networks in minimally twisted bilayer graphene under strain},}\ }\href@noop {} {\bibfield  {journal} {\bibinfo  {journal} {The Journal of Physical Chemistry C}\ }\textbf {\bibinfo {volume} {124}},\ \bibinfo {pages} {2120--2130} (\bibinfo {year} {2019}{\natexlab{b}})}\BibitemShut {NoStop}%
\bibitem [{\citenamefont {Wang}\ and\ \citenamefont {Hsu}(2024)}]{wang2024electrically}%
  \BibitemOpen
  \bibfield  {author} {\bibinfo {author} {\bibfnamefont {H.-C.}\ \bibnamefont {Wang}}\ and\ \bibinfo {author} {\bibfnamefont {C.-H.}\ \bibnamefont {Hsu}},\ }\bibfield  {title} {\enquote {\bibinfo {title} {Electrically tunable correlated domain wall network in twisted bilayer graphene},}\ }\href@noop {} {\bibfield  {journal} {\bibinfo  {journal} {2D Materials}\ }\textbf {\bibinfo {volume} {11}},\ \bibinfo {pages} {035007} (\bibinfo {year} {2024})}\BibitemShut {NoStop}%
\bibitem [{\citenamefont {Kaliteevski}, \citenamefont {Enaldiev},\ and\ \citenamefont {Fal’ko}(2023)}]{kaliteevski2023twirling}%
  \BibitemOpen
  \bibfield  {author} {\bibinfo {author} {\bibfnamefont {M.~A.}\ \bibnamefont {Kaliteevski}}, \bibinfo {author} {\bibfnamefont {V.}~\bibnamefont {Enaldiev}}, \ and\ \bibinfo {author} {\bibfnamefont {V.~I.}\ \bibnamefont {Fal’ko}},\ }\bibfield  {title} {\enquote {\bibinfo {title} {Twirling and spontaneous symmetry breaking of domain wall networks in lattice-reconstructed heterostructures of two-dimensional materials},}\ }\href@noop {} {\bibfield  {journal} {\bibinfo  {journal} {Nano Letters}\ }\textbf {\bibinfo {volume} {23}},\ \bibinfo {pages} {8875--8880} (\bibinfo {year} {2023})}\BibitemShut {NoStop}%
\bibitem [{\citenamefont {Cazeaux}\ \emph {et~al.}(2023)\citenamefont {Cazeaux}, \citenamefont {Clark}, \citenamefont {Engelke}, \citenamefont {Kim},\ and\ \citenamefont {Luskin}}]{cazeaux2023relaxation}%
  \BibitemOpen
  \bibfield  {author} {\bibinfo {author} {\bibfnamefont {P.}~\bibnamefont {Cazeaux}}, \bibinfo {author} {\bibfnamefont {D.}~\bibnamefont {Clark}}, \bibinfo {author} {\bibfnamefont {R.}~\bibnamefont {Engelke}}, \bibinfo {author} {\bibfnamefont {P.}~\bibnamefont {Kim}}, \ and\ \bibinfo {author} {\bibfnamefont {M.}~\bibnamefont {Luskin}},\ }\bibfield  {title} {\enquote {\bibinfo {title} {Relaxation and domain wall structure of bilayer moir{\'e} systems},}\ }\href@noop {} {\bibfield  {journal} {\bibinfo  {journal} {Journal of Elasticity}\ }\textbf {\bibinfo {volume} {154}},\ \bibinfo {pages} {443--466} (\bibinfo {year} {2023})}\BibitemShut {NoStop}%
\bibitem [{\citenamefont {Park}\ \emph {et~al.}(2025)\citenamefont {Park}, \citenamefont {Park}, \citenamefont {Yananose}, \citenamefont {Ko}, \citenamefont {Kim}, \citenamefont {Engelke}, \citenamefont {Zhang}, \citenamefont {Davydov}, \citenamefont {Green}, \citenamefont {Kim} \emph {et~al.}}]{park2025unconventional}%
  \BibitemOpen
  \bibfield  {author} {\bibinfo {author} {\bibfnamefont {D.}~\bibnamefont {Park}}, \bibinfo {author} {\bibfnamefont {C.}~\bibnamefont {Park}}, \bibinfo {author} {\bibfnamefont {K.}~\bibnamefont {Yananose}}, \bibinfo {author} {\bibfnamefont {E.}~\bibnamefont {Ko}}, \bibinfo {author} {\bibfnamefont {B.}~\bibnamefont {Kim}}, \bibinfo {author} {\bibfnamefont {R.}~\bibnamefont {Engelke}}, \bibinfo {author} {\bibfnamefont {X.}~\bibnamefont {Zhang}}, \bibinfo {author} {\bibfnamefont {K.}~\bibnamefont {Davydov}}, \bibinfo {author} {\bibfnamefont {M.}~\bibnamefont {Green}}, \bibinfo {author} {\bibfnamefont {H.-M.}\ \bibnamefont {Kim}},  \emph {et~al.},\ }\bibfield  {title} {\enquote {\bibinfo {title} {Unconventional domain tessellations in moir{\'e}-of-moir{\'e} lattices},}\ }\href@noop {} {\bibfield  {journal} {\bibinfo  {journal} {Nature}\ ,\ \bibinfo {pages} {1--8}} (\bibinfo {year} {2025})}\BibitemShut {NoStop}%
\end{thebibliography}
%

\end{document}